\documentclass[preprint,12pt]{elsarticle}

\usepackage{graphicx}

\usepackage{amssymb}

\usepackage{lineno}

\usepackage{amsmath}
\usepackage{float}
\usepackage{color}
\usepackage{subcaption}
\usepackage{bm}
\usepackage{amsmath,amsthm,amssymb}
\usepackage{tikz}
\usepackage[colorinlistoftodos]{todonotes}
\usepackage{amsmath}
\usepackage{amsthm}
\usepackage{amsmath}
\usepackage{amssymb}
\usepackage{stmaryrd}
\usepackage{mathtools}
\usepackage{csquotes}
\usepackage{mathtools,calc}
\usepackage{pgfplots}
\usepackage{xfrac}
\usepackage{framed}
\usepackage[ruled,vlined]{algorithm2e}
\usepackage{adjustbox}
\usepackage{makecell}
\usepackage{multirow}
\usepackage{bbding}
\usepackage{ragged2e}
\usepackage{blindtext}

\usepackage{hyphenat}

\DeclareMathAlphabet{\mathbcal}{OMS}{cmsy}{b}{n}

\usepackage[T2A,T1]{fontenc}

\journal{CMAME}

\begin{document}

\justifying

\begin{frontmatter}


\title{Jump penalty stabilisation techniques for under-resolved turbulence in discontinuous Galerkin schemes}



\author[upm]{Jiaqing Kou\corref{cor1}}
\cortext[cor1]{Corresponding author}
\ead{jiaqingkou@gmail.com}
\author[upm]{Oscar A. Marino}
\author[upm,ccs]{Esteban Ferrer}

\address[upm]{ETSIAE-UPM-School of Aeronautics, Universidad Politécnica de Madrid, Plaza Cardenal Cisneros 3, E-28040 Madrid, Spain}
\address[ccs]{Center for Computational Simulation, Universidad Politécnica de Madrid, Campus de Montegancedo, Boadilla del Monte, 28660 Madrid, Spain}


\begin{abstract}
Jump penalty stabilisation techniques have been recently proposed for continuous and discontinuous high order Galerkin schemes \cite{ferrer2017interior,moura2022gradient,MOURA2022114200}. The stabilisation relies on the gradient or solution discontinuity at element interfaces to incorporate localised numerical diffusion in the numerical scheme. This diffusion acts as an implicit subgrid model and stablises under-resolved turbulent simulations. 

This paper investigates the effect of jump penalty stabilisation methods (penalising gradient or solution) for stabilisation and improvement of high-order discontinuous Galerkin schemes in turbulent regime.
We analyse these schemes using an eigensolution analysis, a 1D non-linear Burgers equation (mimicking a turbulent cascade) and 3D turbulent Navier-Stokes simulations (Taylor-Green Vortex problem).

We show that the two jump penalty stabilisation techniques can stabilise under-resolved simulations thanks to the improved dispersion-dissipation characteristics (when compared to non-penalised schemes) and provide accurate results for turbulent flows. The numerical results indicate that the proposed jump penalty stabilise under-resolved simulations and improve the simulations, when compared to the original unpenalised scheme and to classic explicit subgrid models (Smagorisnky and Vreman).
\end{abstract}


\end{frontmatter}

\tableofcontents


\section{Introduction}
High-order methods are attracting attention, since they have shown favou\hyp{rable} properties including improved accuracy, controlled numerical dissipation, and high efficiency in modern architectures, only to mention a few \cite{wang2007high,wang2013high}. Modelling and simulating turbulent flows remains a challenge for the computational fluid dynamics (CFD) community, and high- order methods are no exception. To balance accuracy and computational cost, a finite range of scales need to be resolved by the scheme, while finest (under-resolved) scales need to be dissipated to avoid energy accumulation.  This balance requires to stabilise under-resolved simulation and dissipate the energy of under-resolved scales, where implicit or explicit LES techniques can be used. While implicit LES \cite{hickel2006adaptive,grinstein2007implicit} (iLES) relies on the numerics to provide dissipation at under-resolved scales, explicit LES schemes (e.g., Smagorinsky \cite{Smagorinsky}, Vreman \cite{vreman}) include a physically based dissipative term. Generally speaking, the former relies on numerical analysis to include dissipation, while the latter favours physical arguments. In any case, it is important to understand and control the numerical errors regardless of the included turbulent dissipating mechanism. Understanding the numerical behaviour of the scheme is therefore essential when proposing jump penalty stabilisation for under-resolved turbulence. The resulting implicit or explicit numerical scheme needs to damp waves with high wavenumbers (small scales) while correctly capturing low- to medium-wavenumber waves (large scales).
In this text, we consider jump penalty stabilisation as implicit LES, but we will compare our proposed jump penalty stabilisation results with classic explicit subgrid models. 

A variety of high-order methods are available, including Continuous Gal\hyp{erkin} (CG) \cite{karniadakis2005spectral}, Discontinuous Galerkin (DG) \cite{karniadakis2005spectral,hesthaven2007nodal}, Flux Reconstruction \cite{huynh2007flux}, Spectral Difference \cite{kopriva1996conservative,liu2006spectral}, etc. These methods share the similarity of using high order polynomials to represent the numerical solution inside each mesh element, which enables increased accuracy. The difference between CG and the rest of the methods is that the former imposes $C^0$ continuity across the element interfaces (hence the name continous Galerkin). DG offers controlled dissipation with improved robustness through the use of upwind Riemann fluxes across element interfaces \cite{manzanero2020design}. 
The application of high-order methods to simulate turbulent flows is still a challenge, but promising results have been obtained when considering underresolved direct numerical simulation (or iLES) \cite{https://doi.org/10.1002/fld.3943,moura2017eddy,wang2017towards}.

Several efforts have been made to offer numerical dissipation to enable iLES simulation for high-order methods. Different stabilisation methods have been summarised in \cite{braack2007stabilized,burman2008stabilization}. A natural choice to enhance the robustness and control dissipation for DG is to rely on upwind Riemann solvers to dissipate small-scale flow structures that cannot be resolved by the selected discretisation (including grid size and polynomial order). The stabilising effect of Riemann solvers has been studied using von Neumann analyses \cite{Gassner2013,moura2015linear,alhawwary2018fourier} for advection, showing promise as the solution jump correlates with the underresolved scales. 
Another option is to apply Spectral Vanishing Viscosity (SVV) \cite{maday1989analysis,kirby2006stabilisation} originally developed for continuous spectral methods, which includes additional dissipation that varies with the spectral modes that form the solution. It applies the high viscosity effect only to the highest polynomial modes, while maintaining favourable properties for low polynomial modes. SVV is not only useful as a substitute for upwind Riemann solvers, but as a complement to the latter to adjust energy accumulations in medium wavenumbers, which cannot be controlled by the numerical flux \cite{manzanero2020design}. In addition, when SVV is applied to CG, similar behaviour can be observed compared to that obtained using Riemann solvers in DG \cite{moura2016eigensolution}. 

An alternative to linking the stabilising dissipation to the advection fluxes is to consider the diffusive fluxes. Ferrer \cite{ferrer2017interior} proposed novel DG stabilisation for underresolved turbulent flows and the incompressible Navier-Stokes equations, where the penalty parameter included in the DG Interior Penalty (IP) formulation was increased to provide stabilisation. When considering continuous methods like CG ($C^0$ continuity is imposed across the element interfaces), there are no jumps in the solution, but dissipative effects can be introduced considering the jumps of the gradient of the solution. This leads to the Continuous Interior Penalty (CIP) method \cite{douglas1976interior} and the Gradient Jump Penalty (GJP) stabilisation \cite{burman2004edge}, where the penalty is introduced for the jump of the solution gradient at the element interfaces. This method has been shown to be effective in stabilising laminar and turbulent flow simulations in \cite{burman2007continuous} and \cite{moura2022gradient,MOURA2022114200}. An advantage of jump penalty stabilisations is that the stabilisation is independent of the scheme, therefore, no change is needed for the original numerical methods. In addition, this stabilisation strategy is symmetric and completely decoupled from the time discretisation \cite{moura2022gradient,MOURA2022114200}. Therefore, it is potentially very useful for DG stabilisation, where the $C^0$ continuity is not required (and both gradient or solution jump penalties can be considered).

This study investigates the effect of jump penalty methods (penalising gradient or solution) for stabilisation and improvement of high-order DG schemes in turbulent regime. The remaining part of this paper is organised as follows: The jump penalty stabilisation for gradient and solutions is introduced in Section \ref{sec:numerical}. Eigensolution analyses are then detailed in Section \ref{sec:eigen}. In Section \ref{sec:cases}, two turbulent cases are discussed, including Burgers turbulence and Taylor-Green vortex. Finally, conclusions are included in Section \ref{sec:conclusion}.

\section{Jump penalty stabilisation}
\label{sec:numerical}
For simplicity, we first explain the gradient and jump penalty stabilisation for the 1D advection-diffusion equation, and then detail the implementation for the 3D compressible Navier-Stokes (NS) equations in conservative form. 

\subsection{Jump stabilisation for the 1D advection-diffusion equation}
For simplicity, we define the jumps for one-dimensional problems and then extend it to the compressible NS equations. To explain the basic idea of the jump penalty stabilisation, we use the following advection-diffusion equation: 

\begin{equation}
\label{eq:ade}
    \frac{\partial u}{\partial t} + a \frac{\partial u}{\partial x}= \mu \frac{\partial^2 u}{\partial x^2},
\end{equation}
where $a$ is the advection speed, and $\mu$ is the viscosity. Consider the discretization of DG based on a uniform grid with element $\Omega_n$ and size $h$, DG approximates the solution in each element by a polynomial of order $P$:

\begin{equation}
\label{eq:poly}
    u(x,t) = \sum_{j=0}^{P} \hat{u}_j (t) \phi_j (\xi),
\end{equation}
where $\phi_j$ is the modal basis function based on Legendre polynomials, and $\hat{u}_j$ is the modal coefficient of this mode. The element is transformed into a local reference domain $\xi \in [-1,1]$.  The weak form of the advection-diffusion equation requires first to introduce the auxiliary variable $g = \frac{\partial u}{\partial x}$ (with a derivative operator to the reference space $\frac{\partial u}{\partial x} = \frac{h}{2} \frac{\partial u}{\partial \xi}$). Then the weak form is obtained by projecting the equation onto the test basis $\phi_i$, and 

\begin{equation}
\label{eq:main}
    \int_{\Omega_n}^{} \phi_i \frac{\partial u}{\partial t} dx + a  \int_{\Omega_n}^{} \phi_i \frac{\partial u}{\partial x}  dx = \mu \int_{\Omega_n}^{} \phi_i \frac{\partial^2 u}{\partial x^2} dx,
\end{equation}

\begin{equation}
    \int_{\Omega_n}^{} \phi_i g  dx = \int_{\Omega_n}^{} \phi_i \frac{\partial u}{\partial x} dx.
\end{equation}

Through integration by part and using the orthogonality of the basis functions, one can arrive at

\begin{equation}
    \frac{h}{2} \frac{\partial \hat{u}_i}{\partial t} + (au)^* \phi_i |^{+1}_{-1} - \int_{-1}^{1} a u \frac{\partial \phi_i}{\partial \xi} d\xi  = (\mu g)^*|^{+1}_{-1} - \int_{-1}^{1} \mu g \frac{\partial \phi_i}{\partial \xi} d\xi,
\end{equation}

\begin{equation}
    \frac{h}{2}\int_{-1}^{1} g \phi_i d\xi = u^* \phi_i |^{+1}_{-1} - \int_{-1}^{1} u \frac{\partial \phi_i}{\partial \xi} d\xi,
\end{equation}
where the interface values $(au)^*$ and $(\mu g)^*$ are replaced by the numerical fluxes, which are obtained from solutions in adjacent elements and a proper Riemann solver. $+1$ and $-1$ refer to the right ($\xi = 1$) and left ($\xi = -1$) interfaces, respectively. 


The jump penalty stabilisation for DG can be imposed through penalising the jumps from either the gradient or the solution. The gradient jump penalty includes the following source term on the right-hand side of Equation \ref{eq:main}:

\begin{equation}
    \tau_g a h^2 \left\{ \frac{\partial \phi_i }{\partial x} G_{jump} \right\}^{+1}_{-1},
\end{equation}
and the solution jump penalty adds the following term:

\begin{equation}
\label{eq:jumpSolDefinition}
    \tau_s a \left\{ \phi_i U_{jump} \right\}^{+1}_{-1}.
\end{equation}

The discretization of the gradient jump term is as follows:

\begin{equation}
\label{eq:gradJP}
    \left\{ \frac{\partial \phi_i }{\partial x} G_{jump} \right\}^{+1}_{-1} = \frac{\partial \phi_i}{\partial x}(+1)  G_{n,n+1} - \frac{\partial \phi_i}{\partial x}(-1) G_{n-1,n},
\end{equation}
where $n-1$, $n$, $n+1$ refer to the left, current, and right elements, respectively. The jump terms are written as: 

\begin{equation}
\label{eq:gradJP2}
    G_{n,n+1} = (2/h)\left [ \left( \frac{\partial u }{\partial \xi} \right)_{n+1}(-1) - \left(\frac{\partial u }{\partial \xi }\right)_{n}(+1) \right ], 
\end{equation}

\begin{equation}
\label{eq:gradJP3}
    G_{n-1,n} = (2/h)\left [ \left( \frac{\partial u }{\partial \xi} \right)_{n}(-1) - \left(\frac{\partial u }{\partial \xi }\right)_{n-1}(+1) \right ].
\end{equation}

The discretization of the solution jump term is as follows:
\begin{equation}
\label{eq:SolJP}
    \left\{ \phi_i U_{jump} \right\}^{+1}_{-1} = \phi_i(+1) U_{n,n+1} - \phi_i(-1) U_{n-1,n},
\end{equation}
where the jump terms are written as: 

\begin{equation}
    U_{n,n+1} = u_{n+1}(-1) - u_{n}(+1),
\end{equation}

\begin{equation}
    U_{n-1,n} = u_{n}(-1) - u_{n-1}(+1).
\end{equation}

\subsection{Jump stabilisation for the 3D compressible Navier-Stokes equations}
\label{sec:JP-NS}
Both jump penalty stabilisations are derived here for the 3D compressible non-dimensional Navier-Stokes equations and implemented in our high order discontinuous Galerkin solver HORSES3D \cite{https://doi.org/10.48550/arxiv.2206.09733}. The NS equations in conservative form can be written as

\begin{equation} \label{cons_eqn}
    \mathbf{q}_t + \nabla\cdot(\mathbf{F}_e-\mathbf{F}_v) =  \mathbf{0},
\end{equation}
where $\mathbf{q}$ is the vector of conservative variables $\mathbf{q} = [\rho,\rho u,\rho v,\rho w,\rho e]^T$, $\mathbf{F}_e$ and $\mathbf{F}_v$ are the inviscid and viscous fluxes, respectively. Applying a similar procedure as in the 1D advection-diffusion equation, multiplying by a locally smooth test function $\phi_i$, and applying the Gauss law on the integral of the flux, we can obtain the expression

\begin{equation}
    \int_{\Omega_n}{\mathbf{q}_t \phi_i} + \int_{\partial \Omega_n}{(\mathbf{F}^*_e-\mathbf{F}^*_v)\cdot \mathbf{n} \phi_i} - \int_{\Omega_n}{(\mathbf{F}_e-\mathbf{F}_v)\cdot\nabla\phi_i} =\mathbf{0} ,
\label{eq:NSDG}
\end{equation}
where $\mathbf{n}$ is the normal vector at the element boundaries $\partial\Omega_n$. The discontinuous fluxes at inter-element faces have been replaced by numerical fluxes, $\mathbf{F}^{*}_e$ and $\mathbf{F}^{*}_v$, to obtain a weak form for the equations for each element \cite{Kopriva_2009}.

Nonlinear inviscid and viscous numerical fluxes can be chosen appropriately to control dissipation in the numerical scheme \cite{manzanero2020design}. The former is often solved using a Riemann solver (i.e., Roe in this work), while for the latter we used the Symmetrical Interior Penalty (SIP) method  \cite{doi:10.1137/S0036142901384162,FERRER2011224,FERRER20127037,ferrer2017interior,MANZANERO20181}. The SIP method is used as is the 3D version of the solution jump penalty, being formulated as

\begin{equation}
    \mathbf{{F}}^*_v\cdot\mathbf{n} = \{\!\!\{\mathbf{F}_v\}\!\!\}\cdot\mathbf{n} - \sigma \llbracket \mathbf{q}\rrbracket,
\label{eq:IP_flux}
\end{equation}
where $\{\!\!\{\bullet\}\!\!\}$ and $\llbracket\bullet\rrbracket$ stand for the average and jump operators, defined as,

\begin{equation}
    \{\!\!\{\mathbf{q}\}\!\!\} = \frac{\mathbf{q}^+ + \mathbf{q}^-}{2},~~\llbracket \mathbf{q}\rrbracket = \mathbf{q}^+\mathbf{n}^++\mathbf{q}^-\mathbf{n}^-,
\label{eq:JumpsAverages1}
\end{equation}
and

\begin{equation}
    \{\!\!\{\mathbf{F}\}\!\!\} = \frac{\mathbf{F}^+ + \mathbf{F}^-}{2},~~\llbracket \mathbf{F}\rrbracket = \mathbf{F}^+\cdot\mathbf{n}^++\mathbf{F}^-\cdot\mathbf{n}^-,
\label{eq:JumpsAverages2}
\end{equation}
where the symbols $+$ and $-$ refer to the left and right interfaces, which is the projection of the values of the left and right elements onto the face. An explicit version of the solution jump penalty can be obtained by replacing Equation \ref{eq:IP_flux} into Equation \ref{eq:NSDG} and rearranging

\begin{equation}
    \int_{\Omega_n}{\mathbf{q}_t \phi_i} + \int_{\partial \Omega_n}{\mathbf{F}^*_e\cdot \mathbf{n} \phi_i} - \int_{\Omega_n}{(\mathbf{F}_e-\mathbf{F}_v)\cdot\nabla\phi_i} - \int_{\partial \Omega_n}{\{\!\!\{\mathbf{F}_v\}\!\!\} \cdot\mathbf{n}\phi_i} =  \int_{\partial\Omega_n}{\sigma \llbracket \mathbf{\tilde{q}}\rrbracket\phi_i},
\label{eq:NSJumpSol}
\end{equation}
making clear that the subtraction for the 1D case is replaced by the sum of the contribution of each face of the element. Note that for Navier-Stokes equations, we only penalise the jump of conservative variables in the momentum equations. Therefore, the penalised variable $\mathbf{\tilde{q}}$ becomes:

\begin{equation}
    \mathbf{\tilde{q}} = \mathbf{\Lambda} \mathbf{q},
\end{equation}
where $\mathbf{\Lambda}$ is the matrix to indicate only the jump terms in the momentum equations are penalised: 
\begin{equation}
    \mathbf{\Lambda} = \begin{pmatrix}
    0 & \mathbf{0}^\mathrm{T} & 0  \\
    \mathbf{0} & \mathrm{\mathbf{Id}_{3\times3}} &\mathbf{0} \\
    0 & \mathbf{0}^\mathrm{T} & 0  \\
    \end{pmatrix},
\end{equation}
where $\mathrm{\mathbf{Id}_{3\times3}}$ is the $3\times3$ identity matrix and $\mathbf{0}$ is a column vector with 3 zero entries, with the aim of selecting only the momentum equations.

In addition, the value of $\sigma$, in Eq. \eqref{eq:NSJumpSol}, can be compared with Equation \ref{eq:jumpSolDefinition}, which scales with the advection velocity. For the NS case, it can be scaled either with the local velocity of the element (part of the inviscid flux), or more naturally with the viscosity (the Reynolds number), as is part of the viscous flux, thus the explicit addition to the right-hand side of the discretised NS equation is:

\begin{equation}
    \frac{\tau_s}{Re}\int_{\partial\Omega_n}{\llbracket \mathbf{\tilde{q}}\rrbracket\phi_i}.
\label{eq:RHSJumpSol}
\end{equation}

As for the gradient jump penalty, the general expression that is added to the right-hand side of the NS equation, as proposed in \cite{moura2022gradient}, is in the form of,

\begin{equation}
    -\sigma\Bigg<\frac{\partial\phi_i}{\partial n}G_{jump}\Bigg>
\end{equation}

\noindent where the bracket term, $\big<\big>$, represents the integration over the faces of the element, and $G_{jump}$ is the numerical gradient-jump (the equivalent 3D version of Equations~\ref{eq:gradJP}, ~\ref{eq:gradJP2} and ~\ref{eq:gradJP3}). To obtain a similar form as for the solution jump, we need to consider three steps: first, the fact partial derivative of the normal is equivalent to the dot product of the gradient and the normal vector, $\partial\phi_i / \partial n = \nabla{\phi_i}\cdot\mathbf{n}$; second, the definition of the numerical gradient-jump in a single face, $G_{jump} = \llbracket \mathbf{\nabla{\tilde{q}}}\rrbracket$; and finally we use the same scaling factor as the solution penalty, i.e. the Reynolds number. Replacing, the penalty term can be expressed as 

\begin{equation}
    -\frac{\tau_g h^2 }{Re} \int_{\partial\Omega_n}{\llbracket\mathbf{\nabla{\tilde{q}}}\rrbracket\nabla{\phi_i}\cdot\mathbf{n}},
\label{eq:RHSGradSol}
\end{equation}
where $h$ is the grid size of the face (averaged). An important remark is that both gradients of Equation \ref{eq:RHSGradSol} are with respect to the physical coordinates ($x_i)$ and not with the reference coordinates ($\xi_i$), in a general mapping, the use of the Jacobian of the transformation and the contravariant vectors is needed, see details in \cite{Kopriva_2009}.

\section{Eigensolution analyses}
\label{sec:eigen}
Dispersion-dissipation behaviour is crucial to quantify numerical errors and to evaluate the stability of numerical schemes. These behaviours can be characterised using eigensolution analysis, which quantifies how the amplitude and frequency of a wave-like solution evolves in time. Eigensolution analysis, also known as von Neumann or Fourier analysis, has been widely applied to access different high-order methods \cite{hu1999analysis,van2007dispersion,gassner2011comparison,vincent2011insights,moura2015linear,alhawwary2018fourier,manzanero2018dispersion}. Apply the eigendecomposition to the global discretization matrix, where dispersion and dissipation characteristics are related to the matrix eigenvalues. The dispersion error, which relates to the error in wave advection, is represented by the modified wavelength of the wave-like solution. The dissipation error, which corresponds to the nonphysical wave damping/amplification, is represented by the modified amplitude of the wave-like solution. 

Furthermore, insights from eigensolution analysis can be obtained to design improved numerical schemes for turbulence simulation \cite{moura2016eigensolution,manzanero2020design,solan2021application}. Eigensolution analysis in multidimensions can also be used to evaluate the effect of mesh quality for high-order schemes \cite{trojak2020effect}. Most of these analyses belong to temporal eigensolution analysis, since the focus of such analysis is on the temporal evolution of the solution for periodic boundary conditions. To investigate the evolution in space, spatial eigensolution analysis \cite{mengaldo2018spatial,mengaldo2018spatial2} has recently been proposed for high-order schemes with inflow-outflow boundary conditions. The dispersion-dissipation behaviour can also be obtained purely from simulation data, as shown in a recent work by the authors \cite{kou2022data}. Detailed explanation of the eigensolution analysis is given in \ref{sec:appendEA}. 

\begin{figure*}[htbp]
\begin{subfigure}{.45\textwidth}
		\includegraphics[width=180pt]{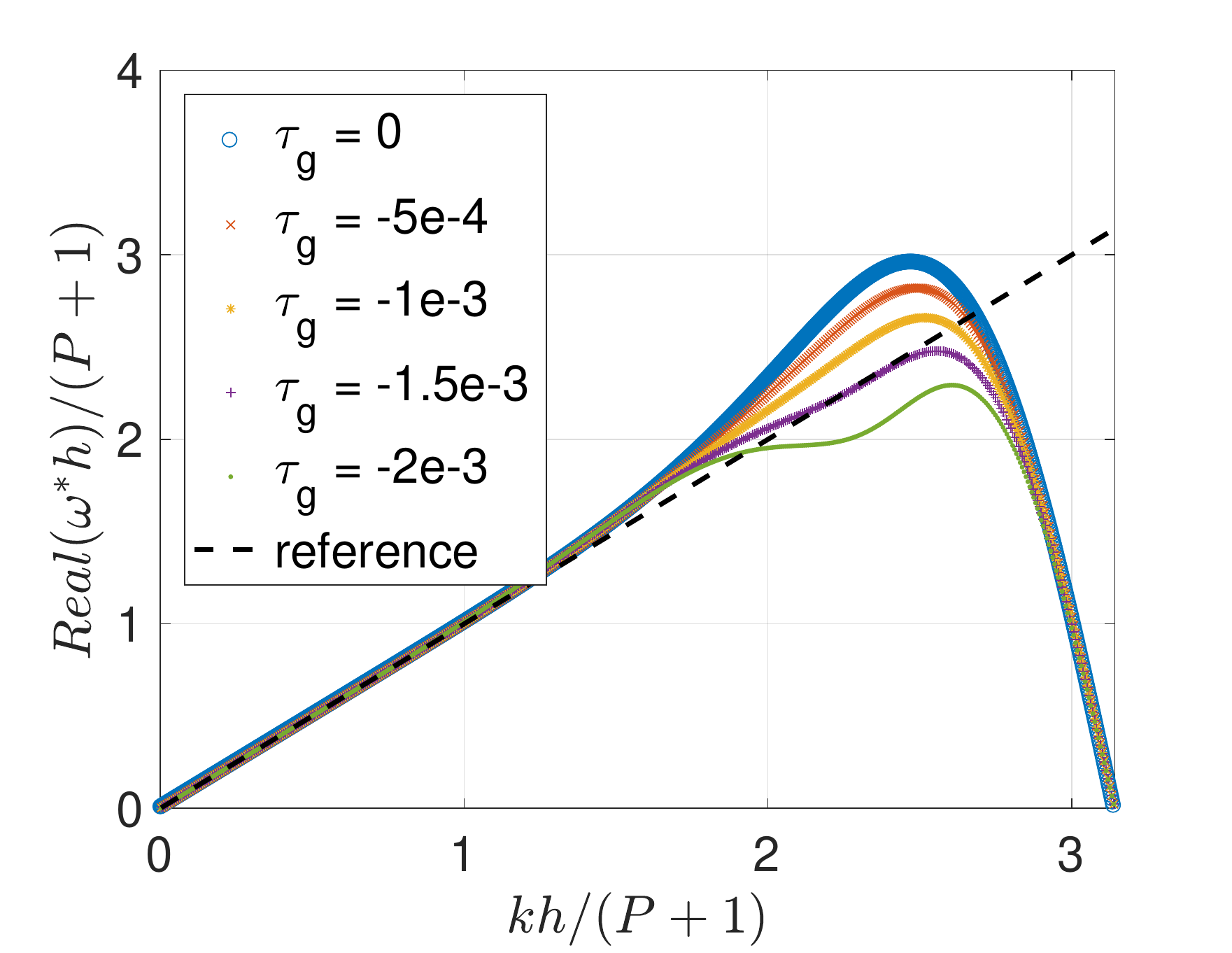}
		\caption{}
	\end{subfigure}
	\begin{subfigure}{.45\textwidth}
		\includegraphics[width=180pt]{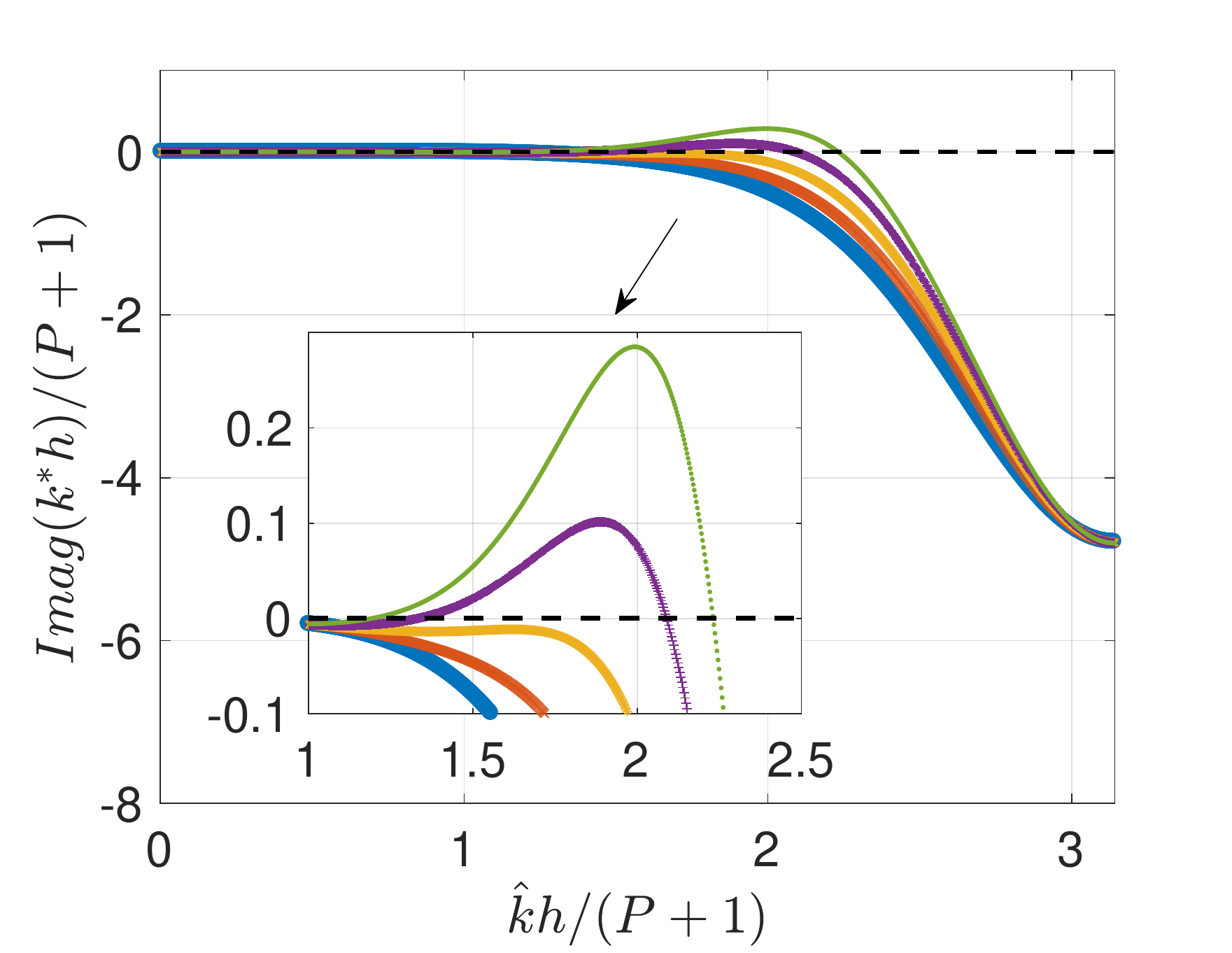}
		\caption{}
	\end{subfigure}
	\centering
	\caption{Eigensolution analysis (gradient jump penalty): Dispersion-dissipation behaviour with $P = 3$ and upwind flux. a) Dispersion. b) Dissipation.}
	\label{fig:P3-grad-up}
\end{figure*}

\begin{figure*}[htbp]
\begin{subfigure}{.45\textwidth}
		\includegraphics[width=180pt]{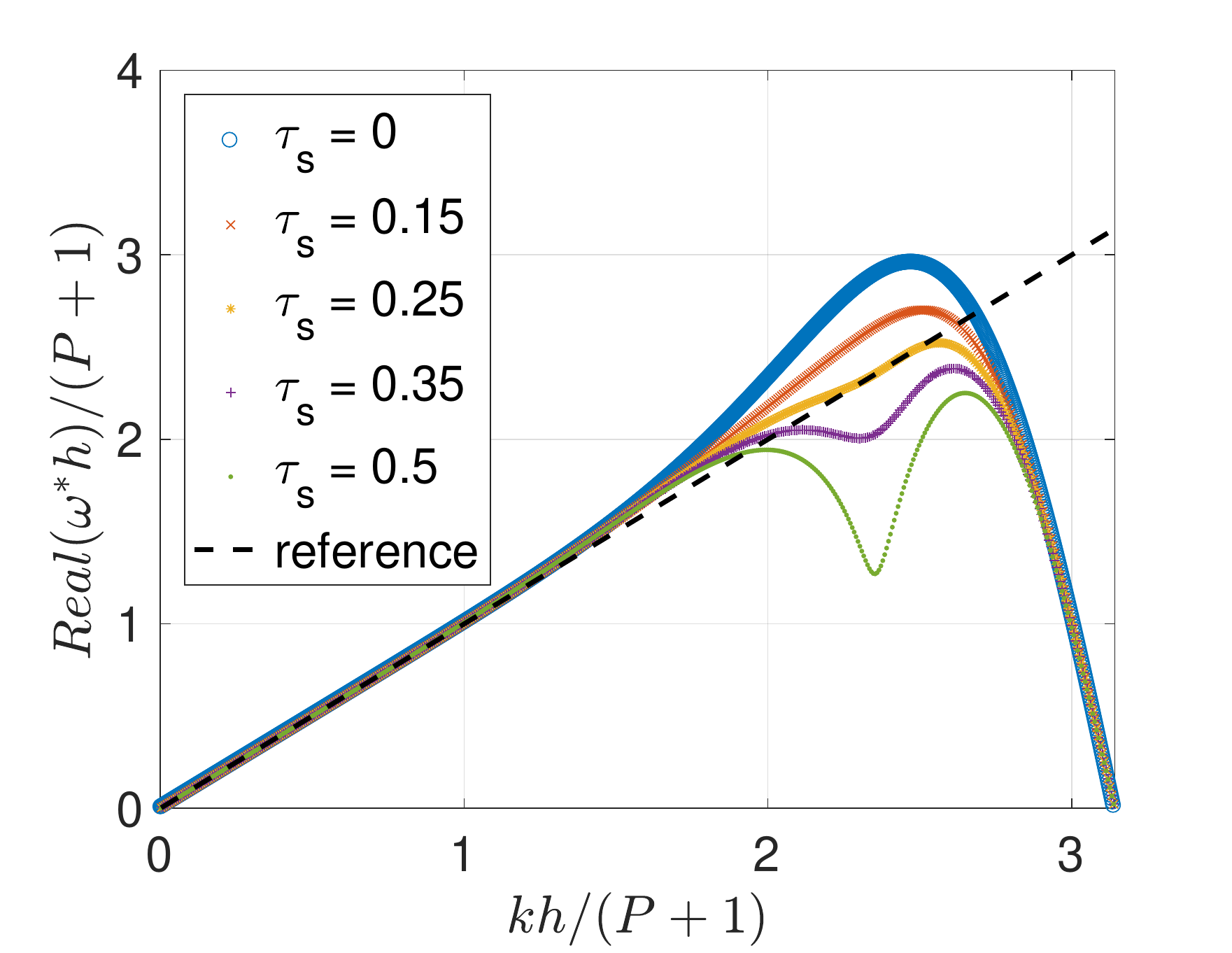}
		\caption{}
	\end{subfigure}
	\begin{subfigure}{.45\textwidth}
		\includegraphics[width=180pt]{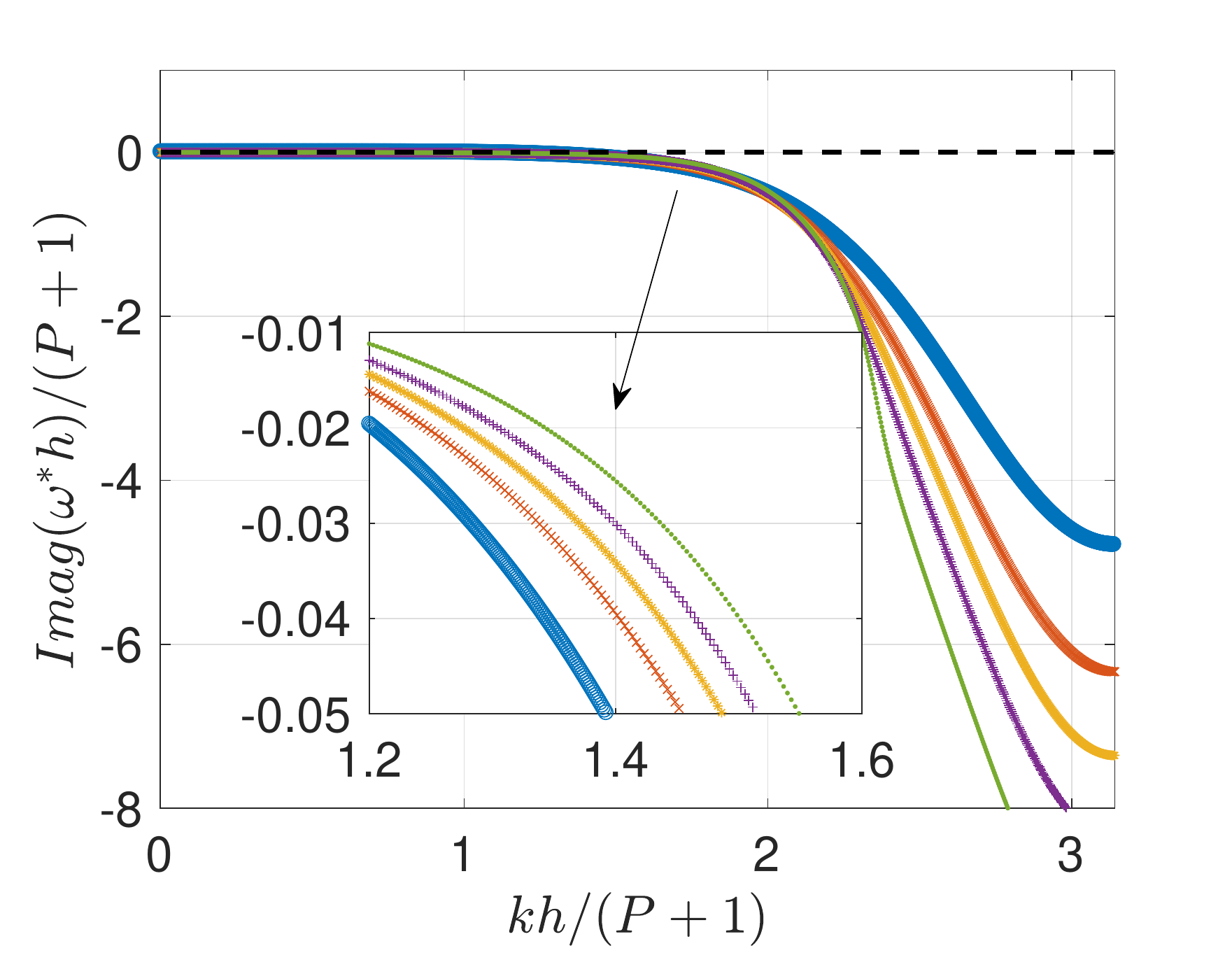}
		\caption{}
	\end{subfigure}
	\centering
	\caption{Eigensolution analysis (solution jump penalty): Dispersion-dissipation behaviour with $P = 3$ and upwind flux. a) Dispersion. b) Dissipation.}
	\label{fig:P3-sol-up}
\end{figure*}

To assess the jump penalty stabilisation, the limit of a very large Péclet number is considered, such that viscous diffusion is negligible. For periodic boundary conditions, the dispersion and dissipation behaviours are investigated by temporal eigensolution analysis. To show the impact of the jump penalty stabilisation, the effect of adding jump penalty terms, as well as the influence of the polynomial order and the type of Riemann solver, are investigated. In all cases we use Gauss-Legendre quadrature. The global matrices needed for the eigensolution analysis are given in \ref{sec:appendDG}. As explained in previous works \citep{moura2015linear,kou2022eigensolution}, a good numerical scheme should have better dispersion property at medium to high wavenumbers, and have more dissipation at high wavenumbers to stabilise the simulation. 

As a representative test case, the influence of the penalty parameters $\tau_g$ and $\tau_s$ is studied for polynomial order $P=3$ and an upwind Riemann flux. The results are compared in Figure \ref{fig:P3-grad-up} and Figure \ref{fig:P3-sol-up}. For standard DG, when an upwind flux is used, the physical mode reflects the true spectral behaviours across all wavenumbers \cite{moura2015linear}. Therefore, only the physical mode is plotted. As shown in Figure \ref{fig:P3-grad-up}, the gradient jump penalty gradually improves the dispersion behaviour when negative $\tau_g$ is used. The dissipation property is also improved, since less dissipation is seen at higher wavenumbers. However, when we continue to decrease $\tau_g$ (increasing the penalty term), e.g., $\tau_g < -1 \times 10^{-3}$, the dissipation becomes positive, indicating that a numerical instability occurs. Therefore, we should limit the gradient jump penalty due to stability reasons, and find the optimal penalty parameter that maintains negative dissipation while optimally improving the dispersion error. As shown in Figure \ref{fig:P3-sol-up}, for the stabilisation of the solution jump penalty, both the dispersion and dissipation behaviours improve, while no numerical instability is induced (no positive dissipation). However, there also exists an optimal penalty parameter that better describes the dispersion behaviour and increases dissipation at high wavenumbers ($\tau_s = 0.25$). The performance of the jump penalty stabilisation for central fluxes, and other polynomial orders (P$=2$ and P$=4$), are included in \ref{sec:app-P2}. From the eigensolution analyses, the optimal penalty parameters for both types of stabilisation are summarised in Table \ref{table2}. As shown in the table, the optimal penalty for the gradient jump stabilisation decreases as the polynomial order increases. Interestingly, the optimal penalty parameter for solution jump stabilisation remains unchanged for all polynomial orders considered.

\begin{figure*}[htbp]
\label{fig:JP-opt-P3}
\begin{subfigure}{.45\textwidth}
		\includegraphics[width=180pt]{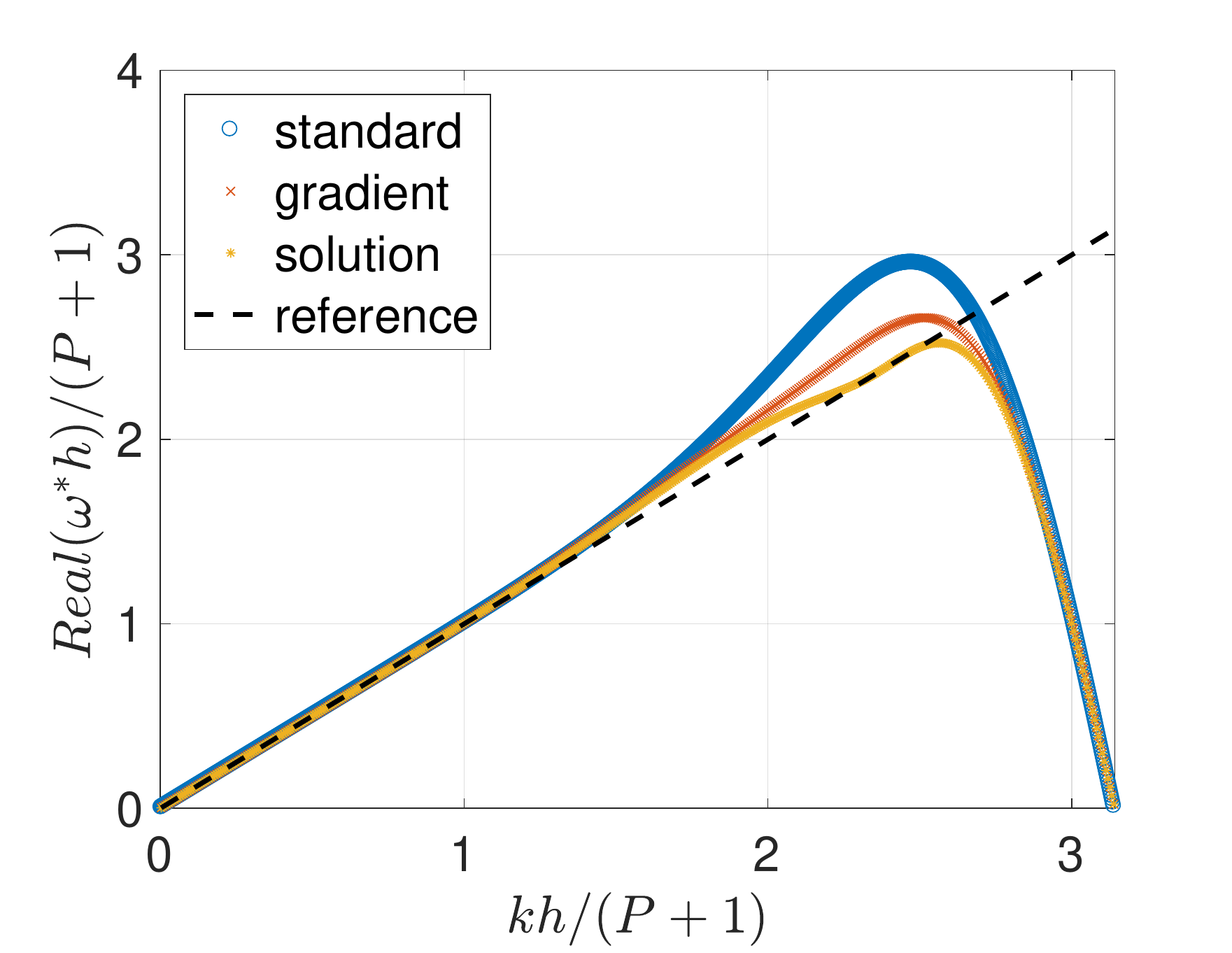}
		\caption{}
	\end{subfigure}
	\begin{subfigure}{.45\textwidth}
		\includegraphics[width=180pt]{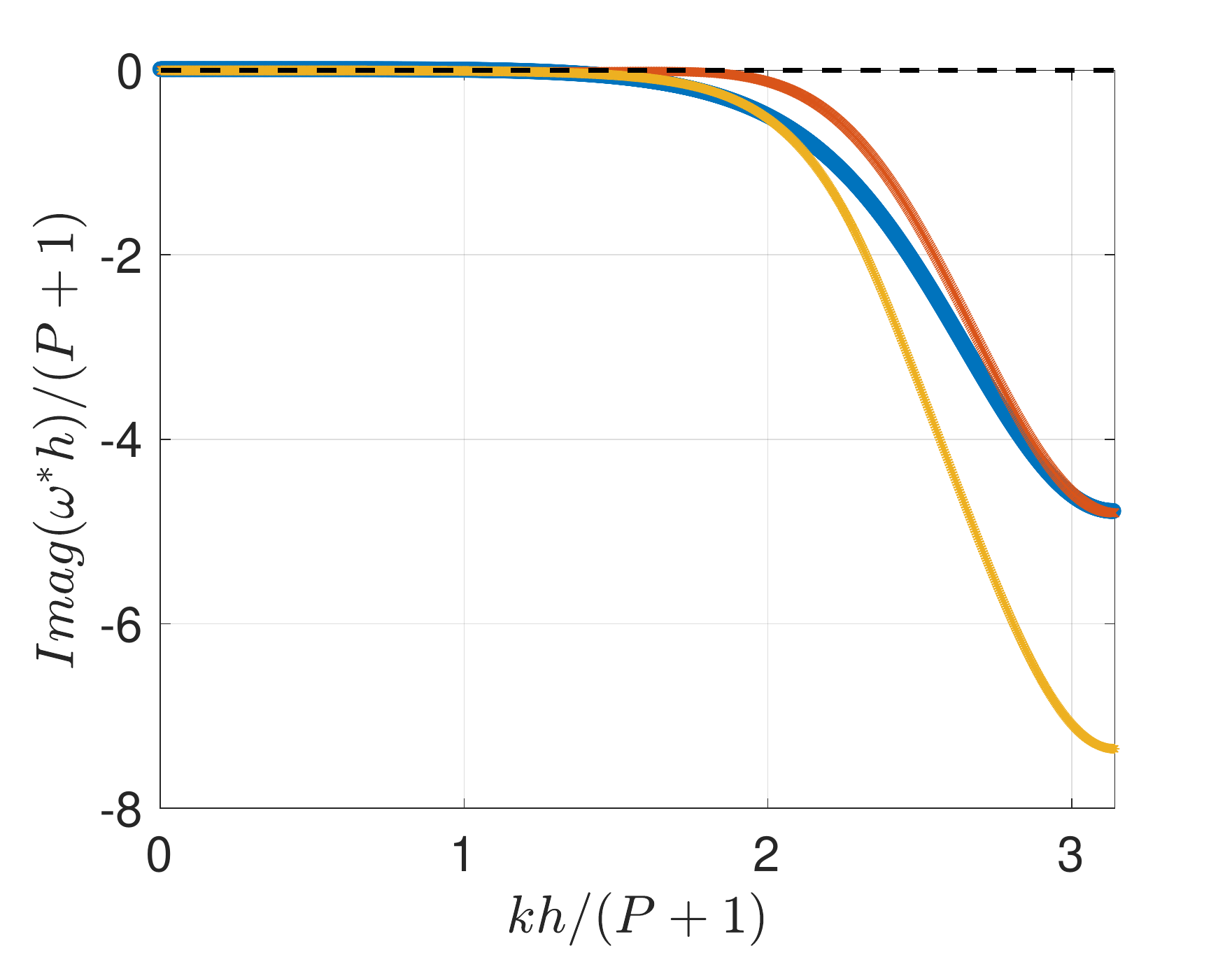}
		\caption{}
	\end{subfigure}
	\centering
	\caption{Eigensolution analysis: Optimal dispersion-dissipation behaviour at P = 3 (upwind flux, $\tau_g = -1\times 10^{-3}, \tau_s = 0.25$). a) Dispersion. b) Dissipation.}
	\label{fig:P3-comp}
\end{figure*}

\begin{table}[htbp!]
	\vspace{20pt}
	\centering
	\begin{tabular}{p{2cm}p{2cm}p{2cm}p{2cm}p{2cm}p{2cm}p{2cm}}
  \hline
 Polynomial & \multicolumn{1}{c}{P = 2} & \multicolumn{1}{c}{P = 3} & \multicolumn{1}{c}{P = 4}\\
  \hline
  Gradient & $-3 \times 10^{-3}$ & $-1 \times 10^{-3}$ & $-4 \times 10^{-4}$ \\
  Solution & $0.25$ & $0.25$ & $0.25$\\
  \hline
 \end{tabular}
	\caption{Eigensolution analysis: Optimal penalty parameters of advection equation for different polynomial orders (upwind Riemann flux).}
	\label{table2}
\end{table}

Figure \ref{fig:P3-comp} compares the optimal spectral behaviours between the two methods of jump penalty stabilisation at polynomial order $P=3$. The optimal curves are obtained through matching the best dispersion behaviour for the jump penalty stabilisation of either gradient or solution, while maintaining negative dissipation to avoid numerical instability. Compared to the standard DG scheme, both the solution and the gradient jump penalty stabilisation improve the spectral properties. In particular, a better dispersion property at medium to high wavenumebers is observed for both methods. For the gradient jump penalty, since numerical instabilities can be induced for large penalties, the penalty parameter should be limited. Therefore, the dispersion behaviour is not as good as that of the solution jump penalty. However, when considering the dissipation property, the gradient jump penalty improves the standard DG scheme by providing minimum dissipation at medium wavenumbers, while the solution jump penalty offers more dissipation for high wavenumbers. These results show the advantages of the jump penalty stabilisation to improve the spectral properties of the standard DG scheme. 

\section{Turbulent cases}
\label{sec:cases}
In this section, the jump penalty stabilisation methods are tested for non-trivial simulations. Two test cases are considered: 1) Forced Burgers turbulence; 2) Taylor-Green vortex. 

\subsection{Burgers Turbulence}
\label{sec:case-BT}
Forced Burgers turbulence \cite{bec2007burgers} has been used to access the spectral behaviour of various DG schemes \cite{moura2015linear,Manzanero2016dispersion}. The governing equation of this problem is as follows:

\begin{equation}
    \frac{\partial u}{\partial t} + \frac{1}{2} \left ( \frac{\partial u^2}{\partial x} \right) = S(x,t),
\end{equation}
where $S$ is the random source term. For Burgers turbulence, a white-in-time random force is used to energise the flow, resulting in a slope of $-5/3$ in the inertial range of the energy spectrum that represents the Navier–Stokes turbulence. This forcing is taken from Manzanero \cite{Manzanero2016dispersion} and Moura et al. \cite{moura2015linear}:

\begin{equation}
    S(x,t) = \frac{h}{2} \frac{A}{\sqrt{\Delta t}} \sum_{n=1}^{Nc} \frac{\sigma_n(t)}{\sqrt{\pi n}} \text{cos}(\frac{2 \pi n}{L} x),
\end{equation}
where $\frac{\Delta x}{2}$ comes from the Jacobian of the element. The cutoff frequency is controlled by $N_c$, which is set to $80$. The amplitude $A$ is set to $0.04$. $\sigma_n(t)$ is the time-dependent random signal following the Gaussian distribution. 

The standard DG scheme based on Gaussian quadrature points (Gauss-Legendre) with Roe flux is considered. The problem is solved in the domain $\Omega \in [-1,1]$ with periodic boundary conditions and is initialised by a constant solution $u_0 = 1$. Time integration is performed based on an explicit third-order Runge–Kutta scheme \cite{gottlieb1998total}. The simulation has been performed until $t = 400$, which is sufficient to reach the statistically steady state. In the test case, the computational domain (periodic domain) is discretised by $N = 400$ elements, and the polynomial order is set to $P = 3$, resulting in $1600$ degrees of freedom (DOFs). Furthermore, the influence of different polynomial orders and a refined mesh has been studied in \ref{sec:appendBT}. The time step is set to ensure $CFL < 0.1$ \cite{moura2015linear}, where $3 \times 10^{-5}$ is selected for all cases.

\begin{figure*}[htbp]
	\begin{subfigure}{1.0\textwidth}
		\includegraphics[width=1.0\textwidth]{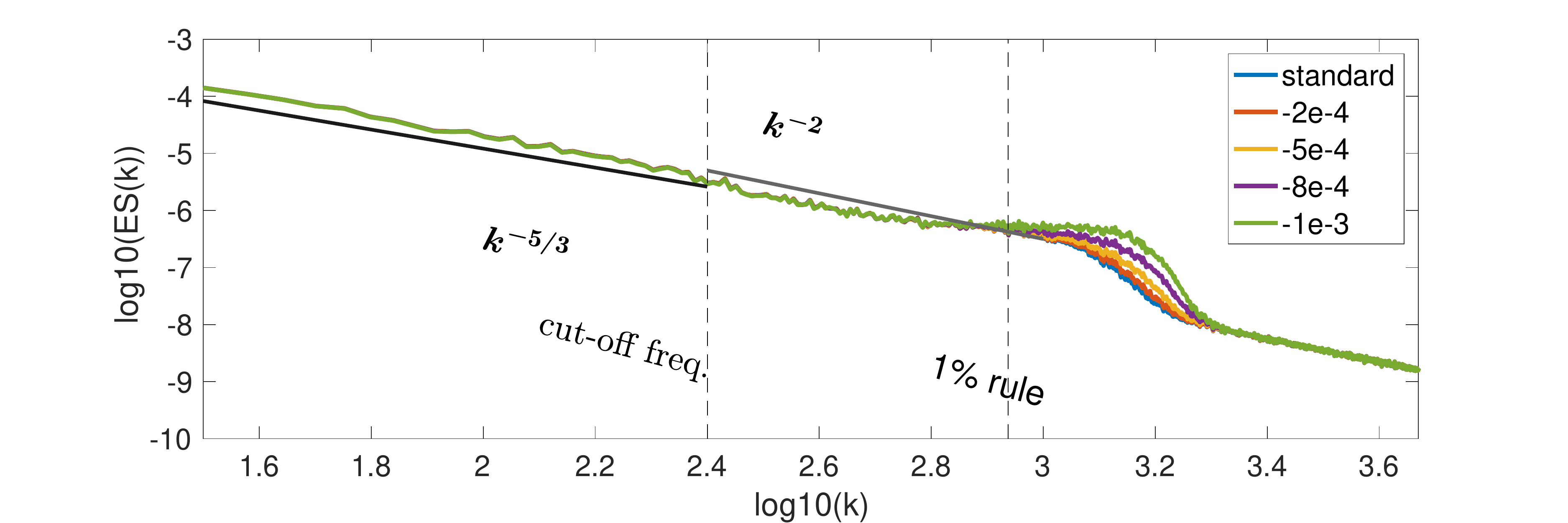}
		\centering
		\caption{Overview.}
	\end{subfigure}
	\begin{subfigure}{1.0\textwidth}
		\includegraphics[width=1.0\textwidth]{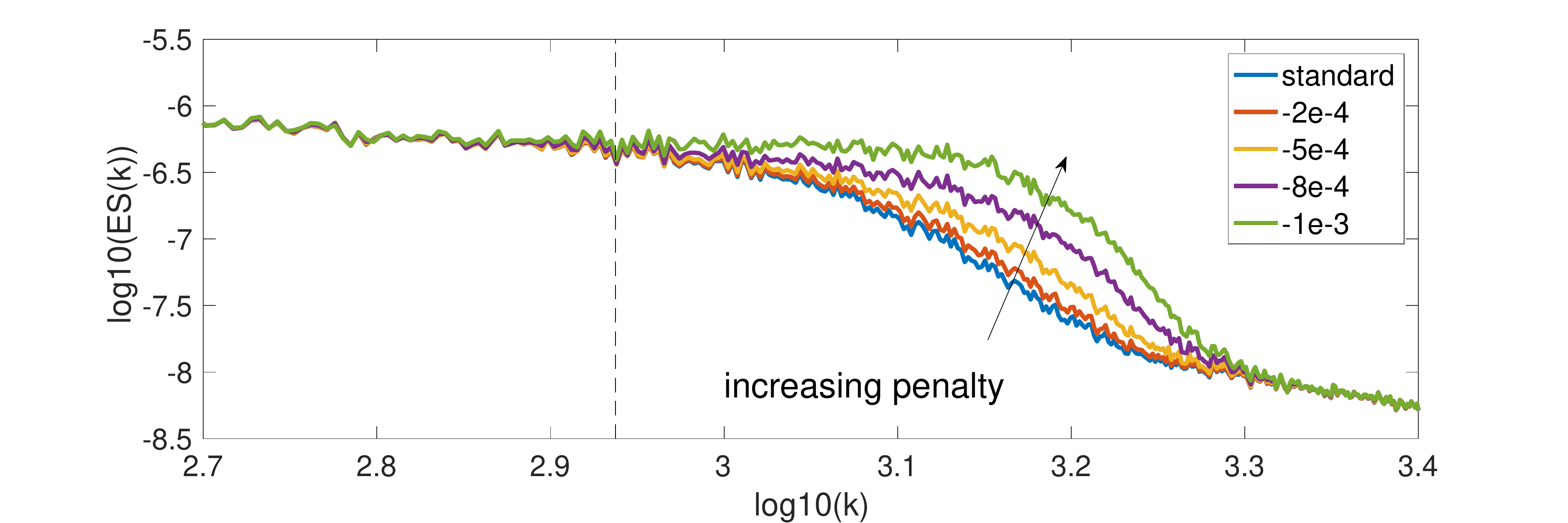}
		\centering
		\caption{Zoom-in view.}
	\end{subfigure}
	\caption{Burgers Turbulence: Time-averaged energy spectrum for the forced Burgers turbulence (gradient jump penalty). The polynomial order is $P = 3$ and the number of element is $N = 400$, resulting in $1600$ total DOFs. This simulation is carried out with Roe fluxes, which are similar to the upwind fluxes considered in the advection equation.}
	\label{fig:Burgers-400g-P3}
\end{figure*}

\begin{figure*}[htbp]
	\begin{subfigure}{1.0\textwidth}
		\includegraphics[width=1.0\textwidth]{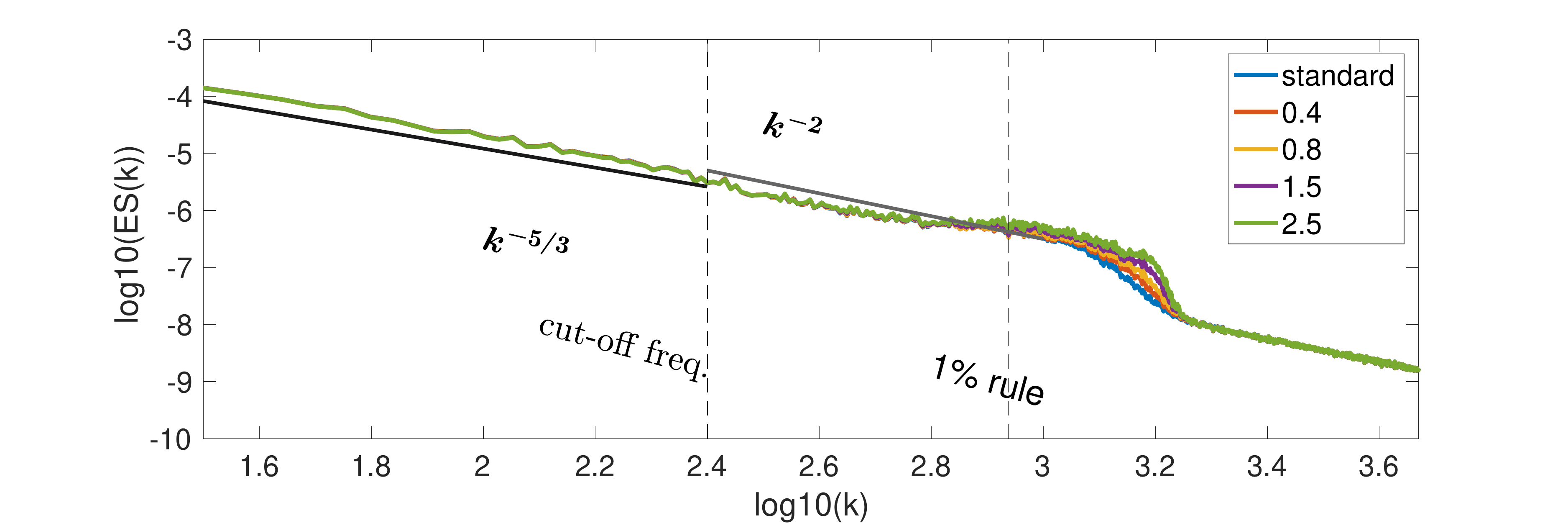}
		\caption{Overview.}
	\end{subfigure}
	\begin{subfigure}{1.0\textwidth}
		\includegraphics[width=1.0\textwidth]{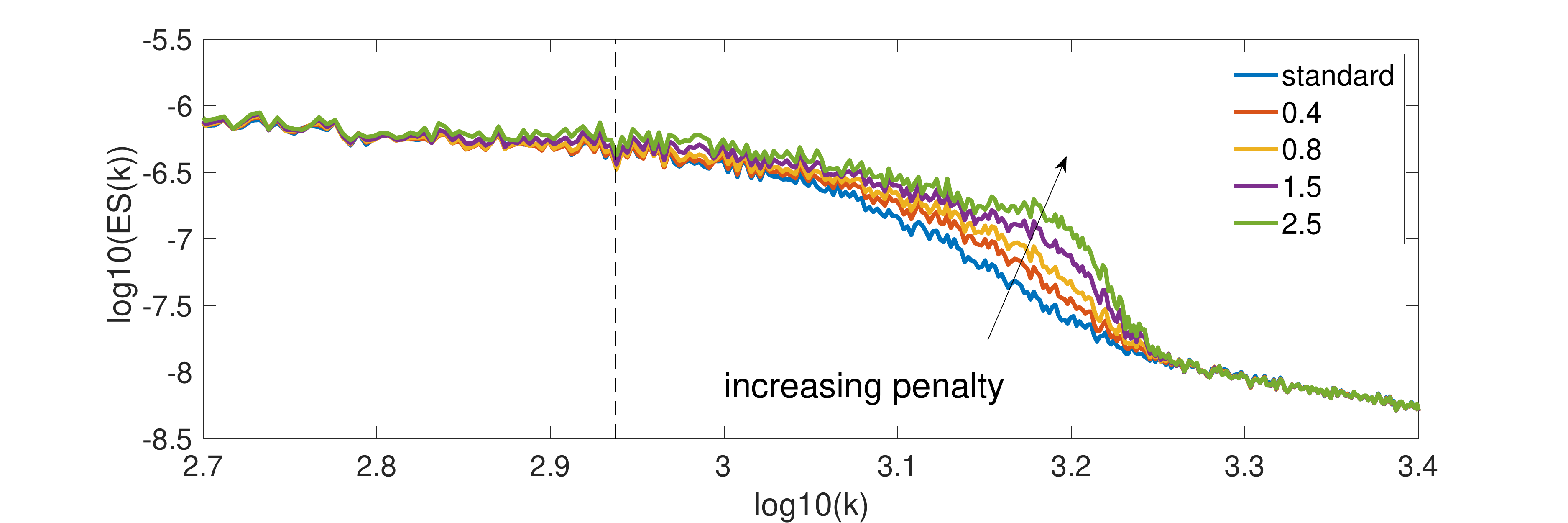}
		\caption{Zoom-in view.}
	\end{subfigure}
	\centering
	\caption{Burgers Turbulence: Time-averaged energy spectrum for the forced Burgers turbulence (solution jump penalty). The polynomial order is $P = 3$ and the number of element is $N = 400$, resulting in $1600$ total DOFs. This simulation is carried out with Roe fluxes, which are similar to the upwind fluxes considered in the advection equation.}
	\label{fig:Burgers-400s-P3}
\end{figure*}

The time-averaged energy spectrum of Burgers turbulence, representing the distribution of turbulent kinetic energy versus wavenumber, is obtained from the temporal evolution of the solution at the probe point $x = 0$ over the time interval $200 \leq t \leq 400$. It should be noted that the average of the solution remains $1$, as the forcing does not affect the $k=0$ component of the solution. Several penalty parameters have been tested for different types of jump penalty stabilisation and the optimal penalty parameter that optimally resolves the turbulent kinetic energy. 

Figures \ref{fig:Burgers-400g-P3} and \ref{fig:Burgers-400s-P3} show the time-averaged energy spectrum for the gradient jump penalty and the solution jump penalty, respectively. A parametric study is shown to indicate the influence of different penalty parameters. The standard results are obtained by numerical simulation based on the standard Roe flux (without penalty). As shown in the figure, the energy spectrum follows the slope of $-5/3$  up to the prescribed forcing frequency $log_{10}(k) = log_{10}(\pi N_c) \approx 2.4$. After that, a typical slope of $-2$ following unforced Burgers turbulence \citep{bec2007burgers} dominates until numerical diffusion dominates to dissipate the small-scale structures. The predicted dissipation range following the $1\%$ rule \citep{moura2015linear} is also highlighted in the figure, which accurately indicates the beginning of numerical dissipation where the slope $-2$ cannot be maintained. 

As shown in Figures \ref{fig:Burgers-400g-P3} and \ref{fig:Burgers-400s-P3}, both types of jump penalty stabilisation improve the performance of under-resolved 1D Burgers turbulence. As the penalisation term is increased, the resolution of the energy spectrum at high wavenumbers is improved, as evidenced by the extended range of the resolved wavenumber. However, when the penalisation term keeps increasing, although more energy is maintained at high wave numbers, the slope of unforced Burgers turbulence ($-2$) cannot be followed. This is not consistent with real flow physics and can lead to numerical instability (due to high energy at high wavenumbers), indicating that there exists an optimal penalty parameter for each stabilisation method. To determine the optimal penalty parameter, we choose $\tau_g$ and $\tau_s$ which maintain the slope of unforced Burgers turbulence ($-2$) while better resolving the turbulence energy spectrum. Therefore, in Figure \ref{fig:Burgers-400g-P3} and Figure \ref{fig:Burgers-400s-P3}, the optimal penalty parameters for both methods are determined as $\tau_g = -8 \times 10^{-4}$ and $\tau_s = 1.5$, respectively. 

\begin{figure*}[htbp]
	\begin{subfigure}{1.0\textwidth}
		\includegraphics[width=1.0\textwidth]{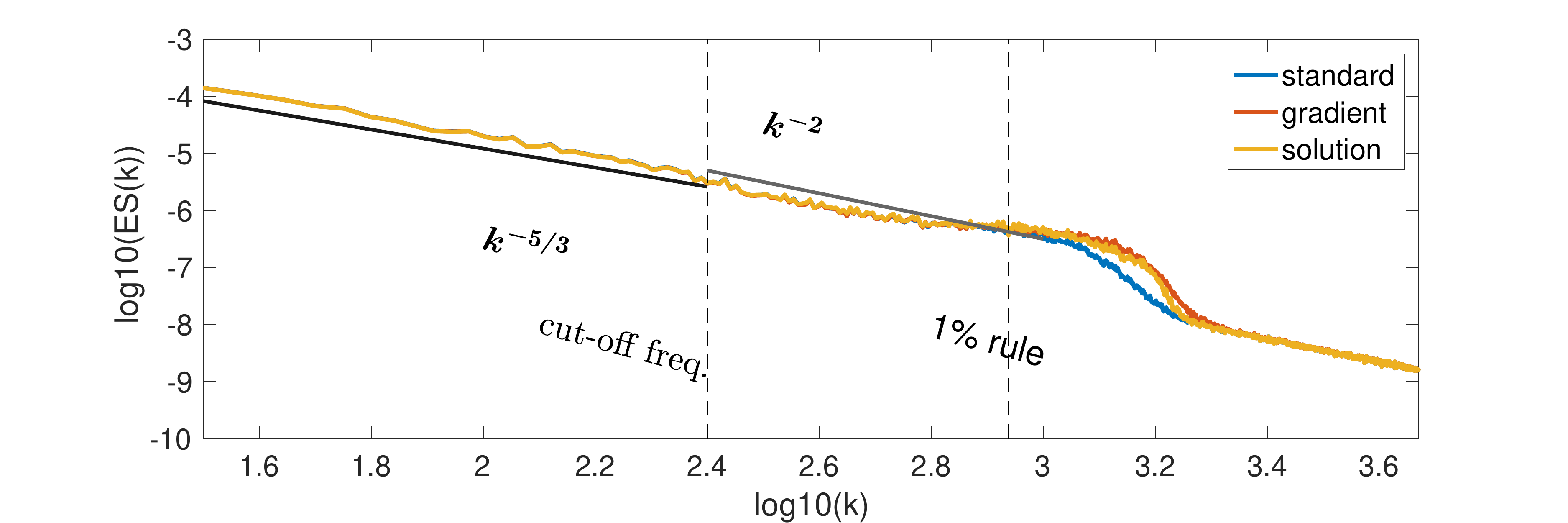}
		\caption{Overview.}
	\end{subfigure}
	\begin{subfigure}{1.0\textwidth}
		\includegraphics[width=1.0\textwidth]{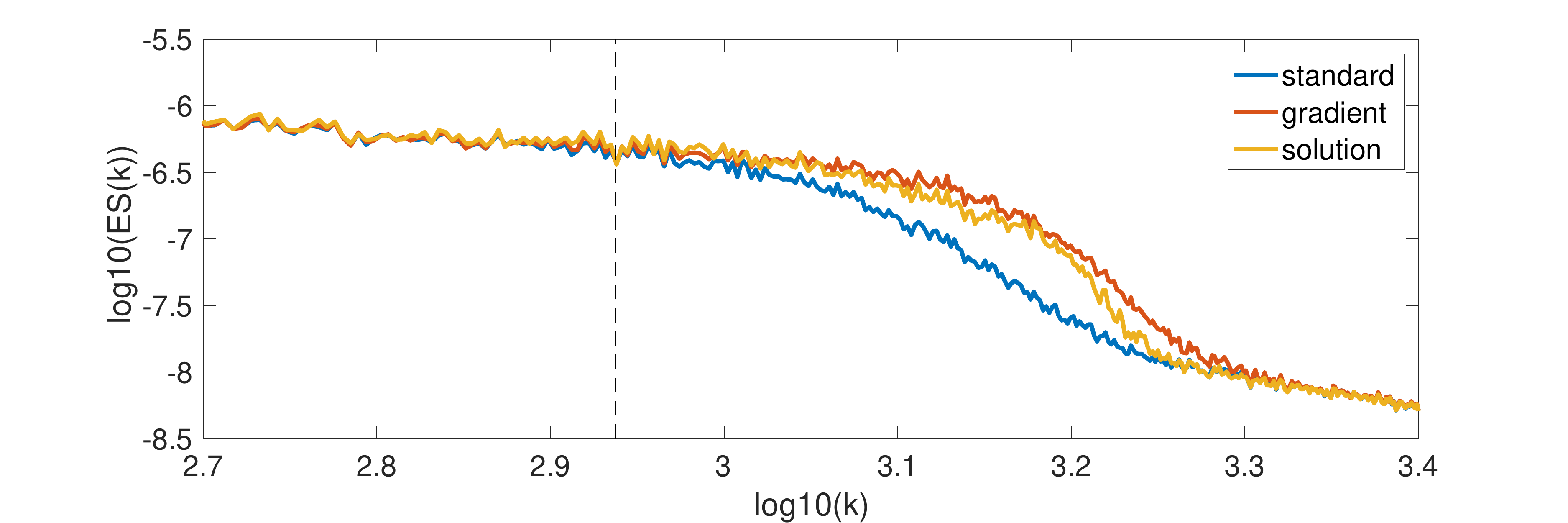}
		\caption{Zoom-in view.}
	\end{subfigure}
	\centering
	\caption{Burgers Turbulence: Time-averaged energy spectrum for the forced Burgers turbulence. The polynomial order is $P = 3$ and the number of element is $N = 400$, resulting in $1600$ total DOFs. This simulation is carried out with Roe fluxes, which are similar to the upwind fluxes considered in the advection equation. (Gradient jump penalty: $\tau_g = -8 \times 10^{-4}$; Solution jump penalty: $\tau_s = 1.5$)}
	\label{fig:Burgers-400-P3}
\end{figure*}

The time-averaged energy spectra for the optimal penalty parameters are compared in Figure \ref{fig:Burgers-400-P3}, where both methods improve the standard DG with similar performance. When the jump penalty stabilisation is used, the dissipation range is further delayed, indicating that the resolved wavenumber range is enlarged. This is similar to adding upwind numerical flux to the numerical scheme, e.g., see Section \ref{sec:appendBT}. This behaviour agrees well with the eigensolution analysis in Section \ref{sec:eigen}, where the jump penalty stabilisation improves the dissipation property by adding less diffusion in the medium wavenumber rage and more diffusion at high wavenumbers. In other words, the jump penalty stabilisation pushes the dissipation curve towards right (shifting to higher wavenumbers) and reduces dissipation for medium wavenumbers. Compared to the solution jump penalty, the gradient jump penalty shows slightly better behaviour than the solution jump, as it maintains the slope $-2$ for higher wavenumbers. Additional results for other polynomial orders $P=2$ and $P=4$, as well as for a refined mesh with $N=819$, are discussed in \ref{sec:appendBT}, where similar behaviours can be observed.

\subsection{Taylor-Green Vortex}

\begin{figure*}
 \centering
  \subfloat[\centering $t/t_c=$ 7.]{
    \includegraphics[width=0.3\textwidth]{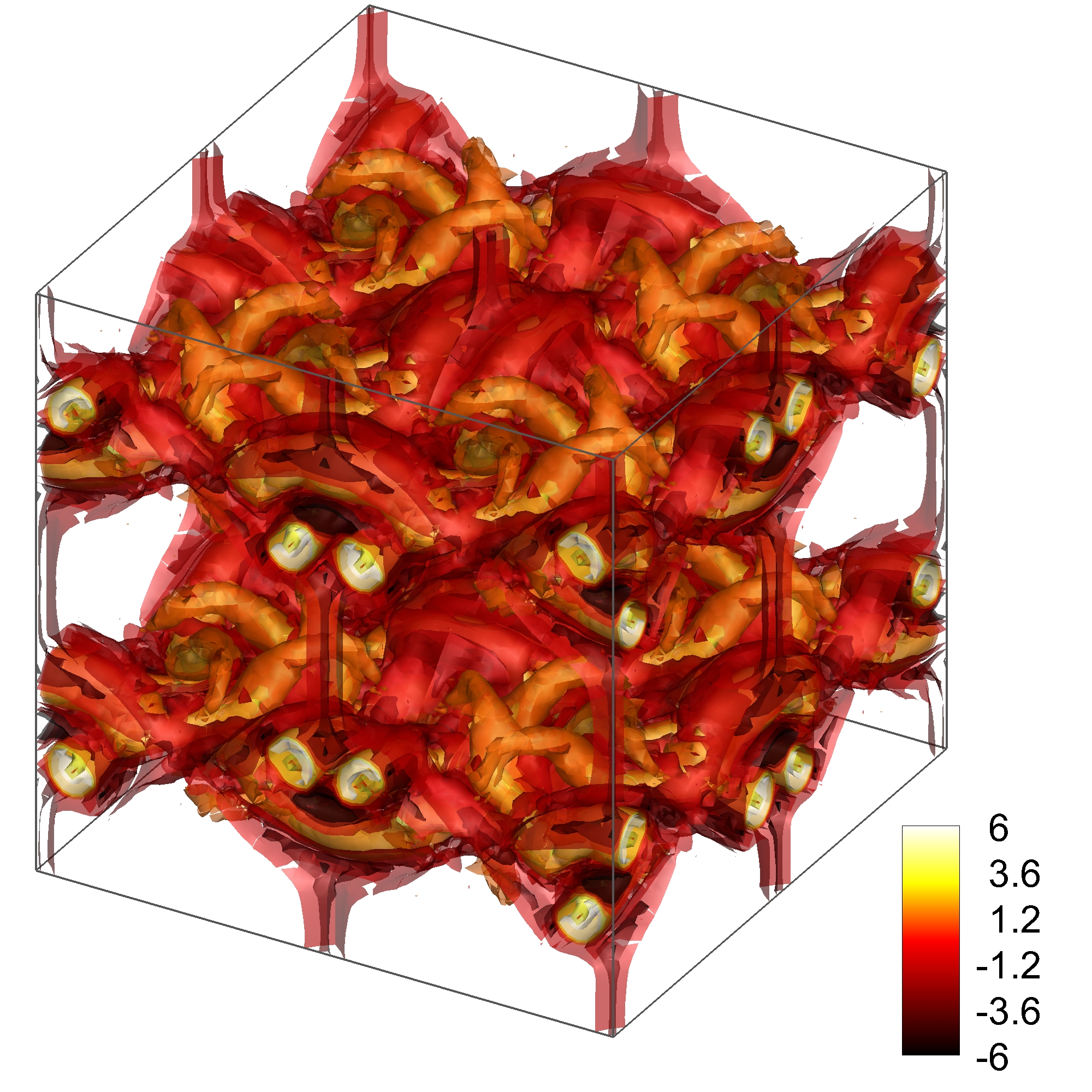}
     \label{Fig:IsoSurfaceLESt7}}
  \subfloat[\centering $t/t_c=$ 10.]{
    \includegraphics[width=0.3\textwidth]{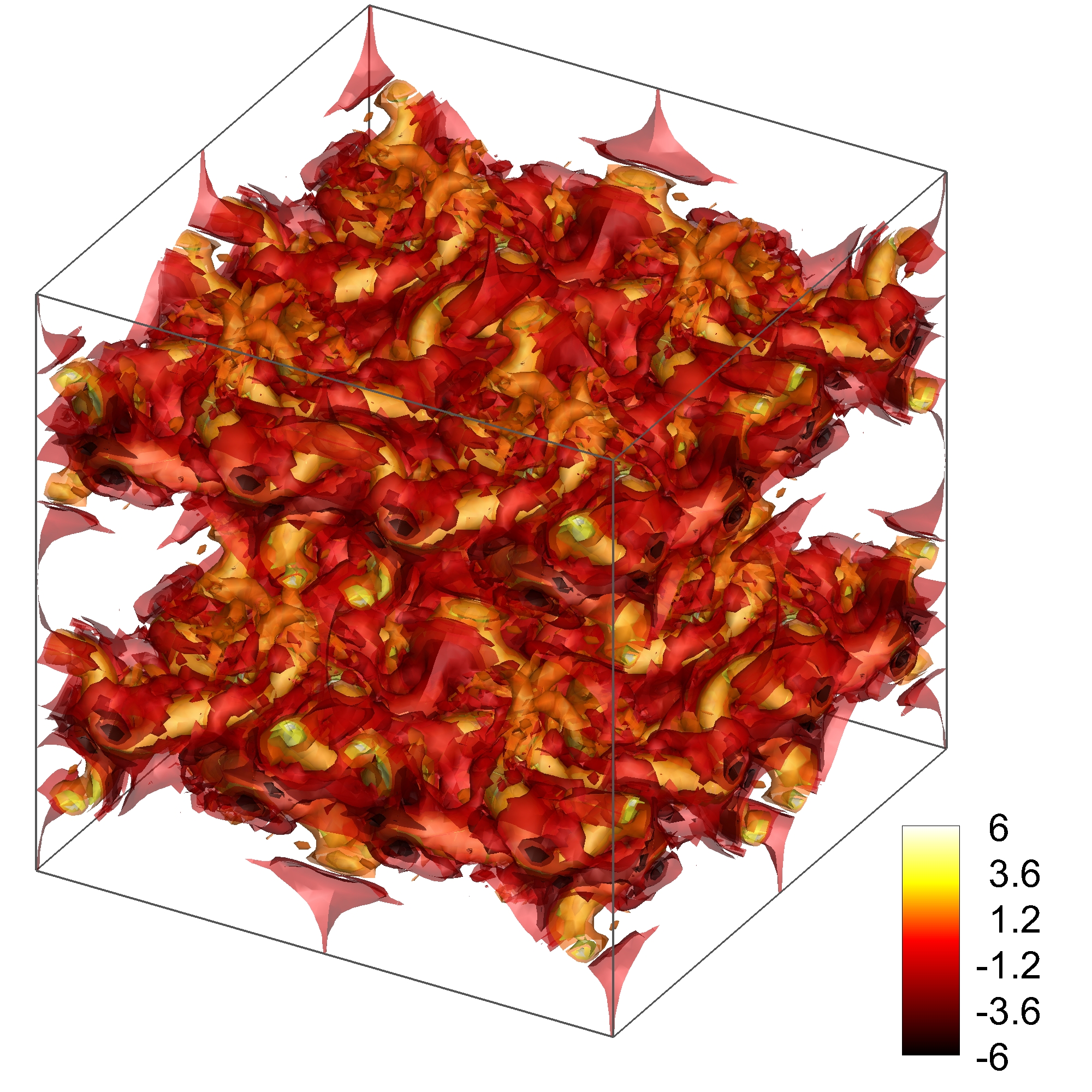}
    \label{Fig:IsoSurfaceLESt10}}
   \subfloat[\centering $t/t_c=$ 13.]{
    \includegraphics[width=0.3\textwidth]{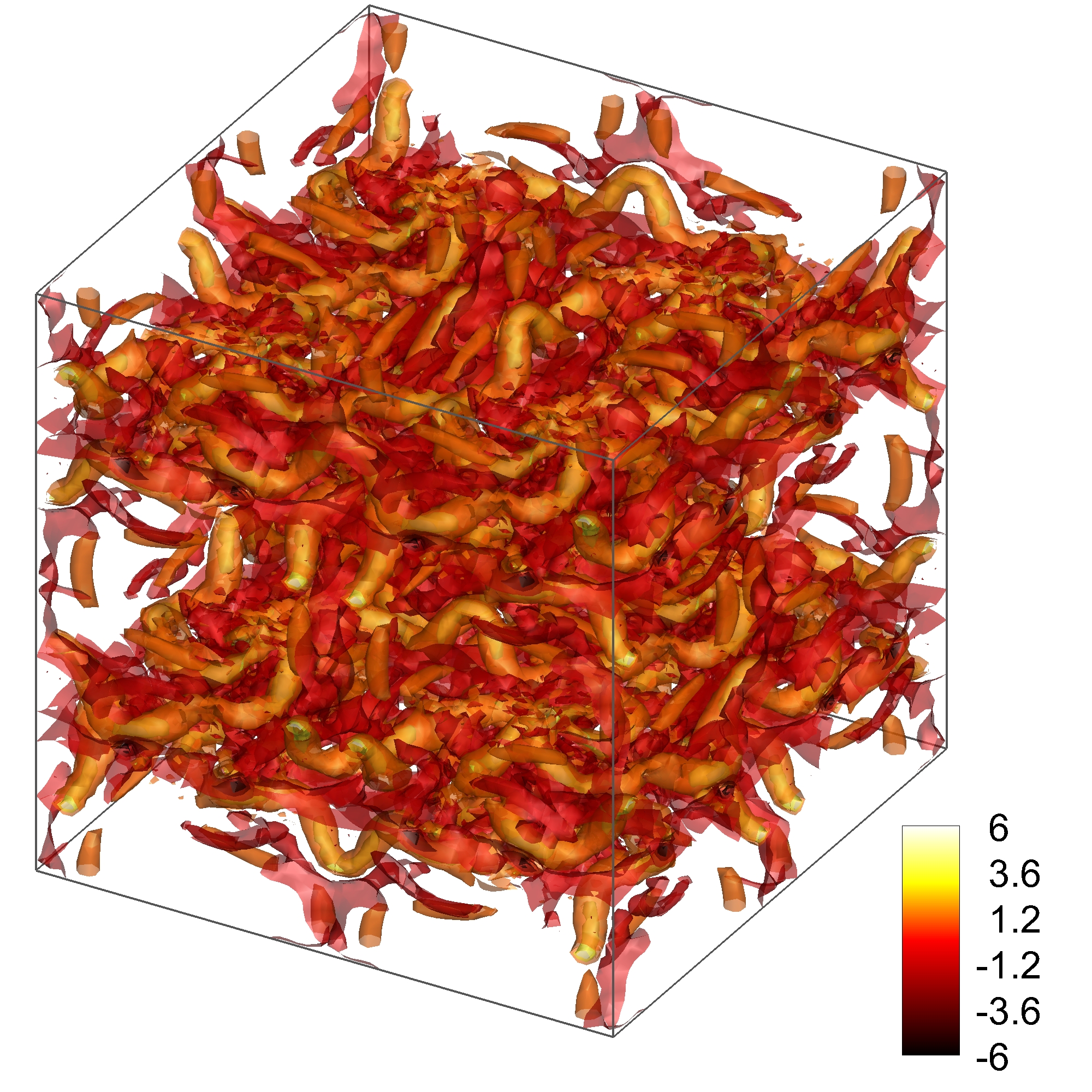}
    \label{Fig:IsoSurfaceLESt13}}
    \caption{Taylor Green vortex: Q-criterion for $Re=$1600 at different times $t/t_c=$7, 10 and 13.}
    \label{Fig:LESDifferentTimes}
\end{figure*}

To challenge the jump penalisation stabilisation in a 3D nonlinear case, we use the fully compressible Navier-Stokes equations to solve the classic Taylor-Green Vortex (TGV) problem. The TGV uses a three-dimensional periodic box $[-\pi,\pi]^3$ with the initial condition,

\begin{equation}
\begin{split}
\rho &= \rho_0,\\
v_1 &= V_0 \sin x\cos y \cos z,\\
v_2 &= -V_0\cos x \sin y \cos z,\\
v_3 &= 0,\\
p &= \frac{\rho_0 V_0^2}{\gamma M_0^2} + \frac{\rho_0 V_0^2}{16}(\cos 2x + \cos 2y)(\cos 2z + 2).\\
\end{split}
\label{eq:TGV}
\end{equation}
The conditions for the case are set for a Reynolds number of $Re=1600$ and a Mach number of 0.1, solved using a Cartesian grid made of $32^3$ elements, each of them enriched with a polynomial of order 3. The TGV problem has been widely used to study numerical methods to analyse how they can reproduce different flow regimes: an initial unstable laminar flow, a transitional laminar-turbulent flow, and a fully turbulent flow with isotropic decay; see Figure \ref{Fig:LESDifferentTimes} and further details in \cite{sharma_2019,Gassner_2016,moura2017diffusion,manzanero2020design}.

\begin{figure*}[htbp]
\begin{subfigure}{1.0\textwidth}
		\includegraphics[width=1.0\textwidth]{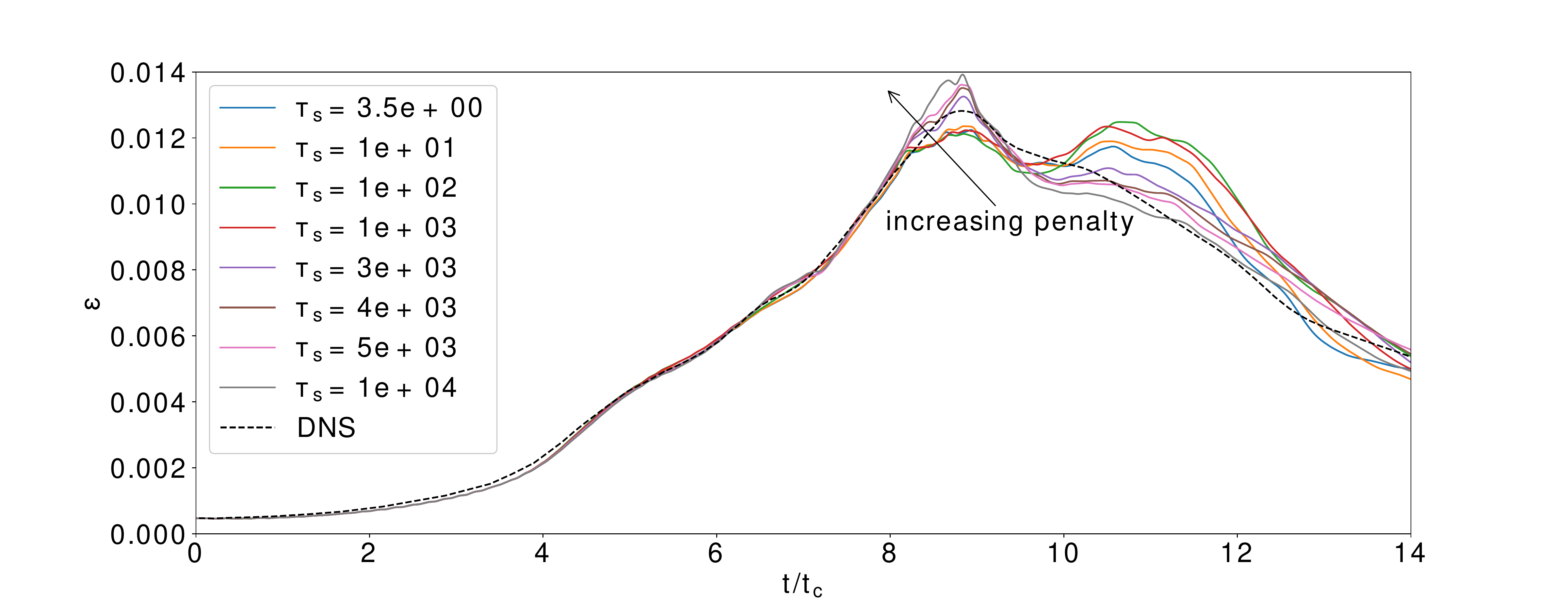}
		\caption{}
	\end{subfigure}
	\begin{subfigure}{1.0\textwidth}
		\includegraphics[width=1.0\textwidth]{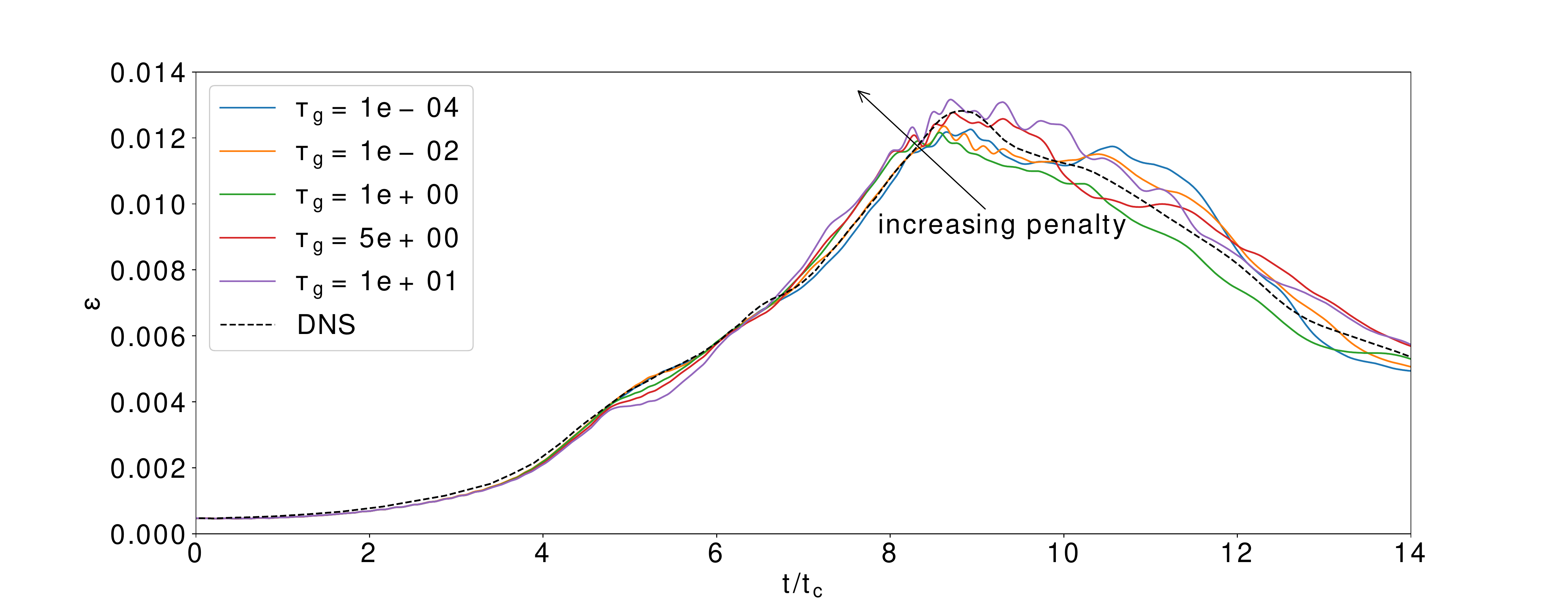}
		\caption{}
	\end{subfigure}
	\centering
	\caption{Taylor Green vortex: Evolution of the kinetic energy dissipation rate in time of the TGV for several penalty parameters compared with DNS results~\cite{wang2013high}. a) Solution Jump. b) Gradient Jump.}
	\label{fig:TGV-dissipation}
\end{figure*}

We compare in Figure \ref{fig:TGV-dissipation} the dissipation rate of kinetic energy as the simulation evolves for the solution and the gradient jump and include a wide range of penalty parameters. Note that if no jump penalty is used, the simulation will fail for the current settings. Figure \ref{fig:TGV-dissipation}a shows that the penalty parameter in the solution jump significantly alters the kinetic energy dissipation rate; in particular, we observe that for low penalties there exist two peaks at $t/t_c=$9 and 11, while the DNS solution shows only one distinct peak at $t/t_c=$9. Increasing the penalty parameters decreases the second peak and recovers a time evolution similar to the DNS. The previous von Neumann analysis suggests that for low penalities there are significant dispersion errors at mid-large wavenumbers that alter the energy dissipation leading to the second artificial peak at $t/t_c=$11. When the penalty parameter is sufficiently large ($\tau_s$ >3000), the second peak is lost and the correct energy dissipation is found. However, we observe a higher first peak (than in the DNS) when increasing the penalty, although the overall shape is correct.  

Figure \ref{fig:TGV-dissipation}b shows the results for the gradient jump penalty stabilisation. Here, we also see that for very low penalisation (baseline configuration) we obtain 2 peaks at $t/t_c=$9 and 11. When the penalty parameter is sufficiently large ($\tau_g$ >1), the second artificial peak is lost, and we recover the DNS shape. In this case, the largest penalty recovers the correct magnitude for the peak at $t/t_c=$9 but provides an oscillatory behaviour at larger times. The overall trend is correct. Note that for both types of jump penalty stabilisation, the penalty parameter is relatively larger than the one used for the forced Burgers turbulence. This is due to the scaling of the penalisation parameter introduced in Section \ref{sec:JP-NS}. Since very large penalisation may be involved, this term can be treated implicitly, which is worth investigating in future works.

\begin{figure*}[htbp]
\begin{subfigure}{.45\textwidth}
		\includegraphics[width=180pt]{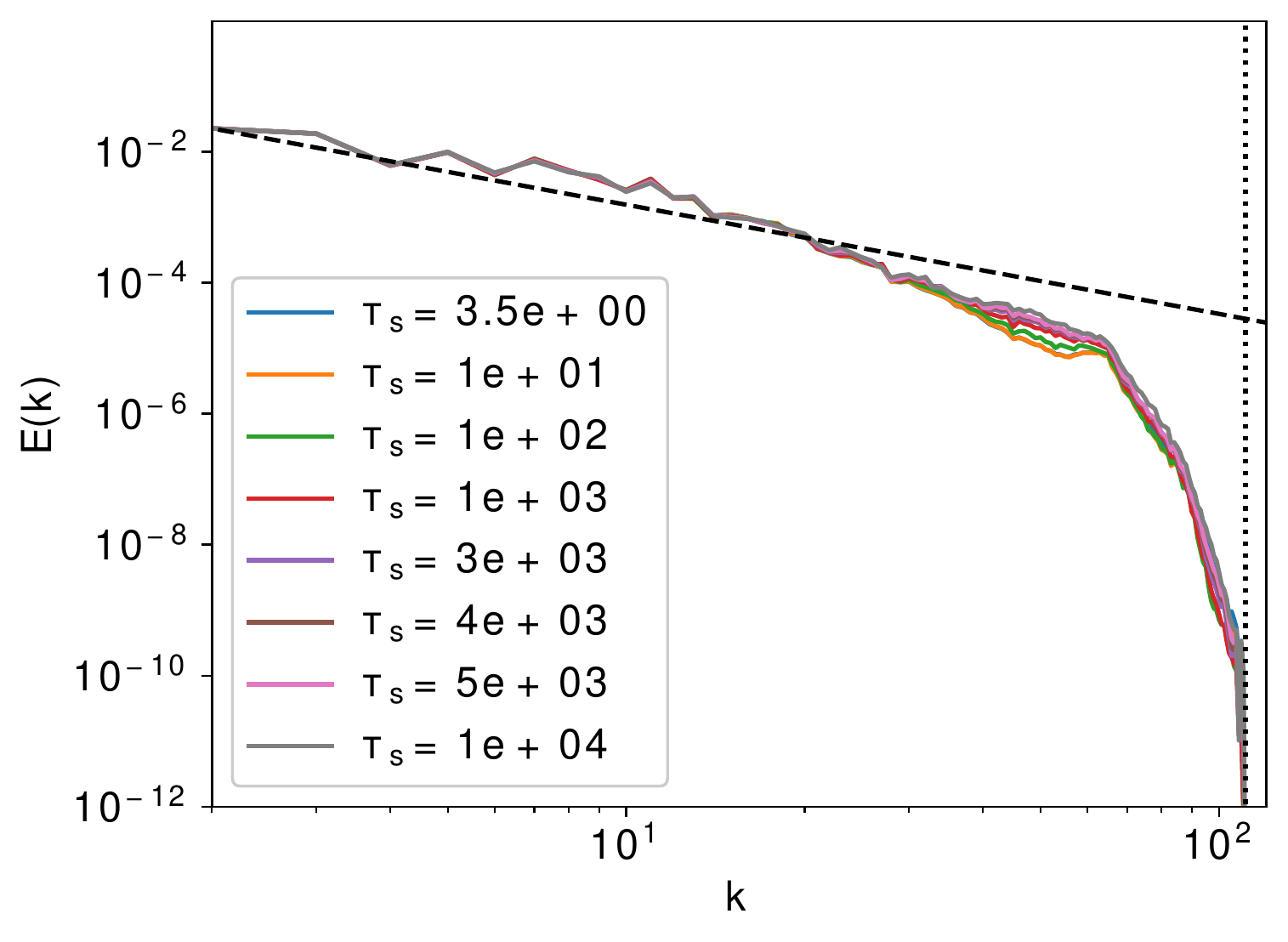}
		\caption{}
	\end{subfigure}
	\begin{subfigure}{.45\textwidth}
		\includegraphics[width=180pt]{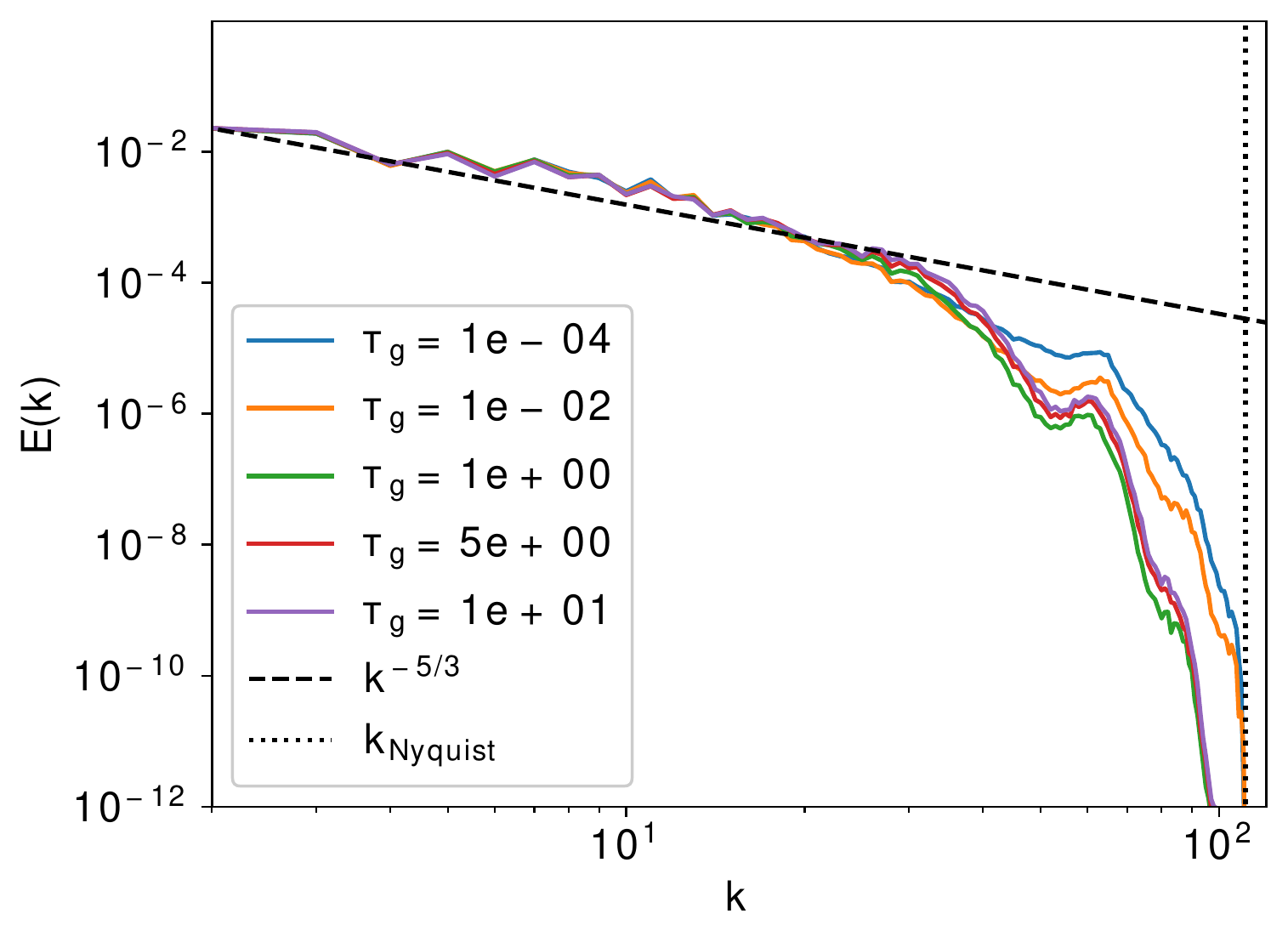}
		\caption{}
	\end{subfigure}
	\centering
	\caption{Taylor Green vortex: Kinetic energy spectra at $t/t_c = 8$ for several penalty parameters. a) Solution Jump. b) Gradient Jump.}
	\label{fig:TGV-dissipation2}
\end{figure*}

Having analysed the integral quantities (over all wavenumbers), we now analyse the energy for all wavenumbers in Figures \ref{fig:TGV-dissipation2}a and b. The Figures depict the kinetic energy spectra at a fixed time $t/t_c = 8$ for the solution and the gradient jump, respectively. We observe that the value of the penalty parameter does not have a significant effect in the case of the solution jump, while the energy decays significantly faster when considering large values for the gradient jump. 

\begin{figure*}[htbp]
\begin{subfigure}{.45\textwidth}
		\includegraphics[width=180pt]{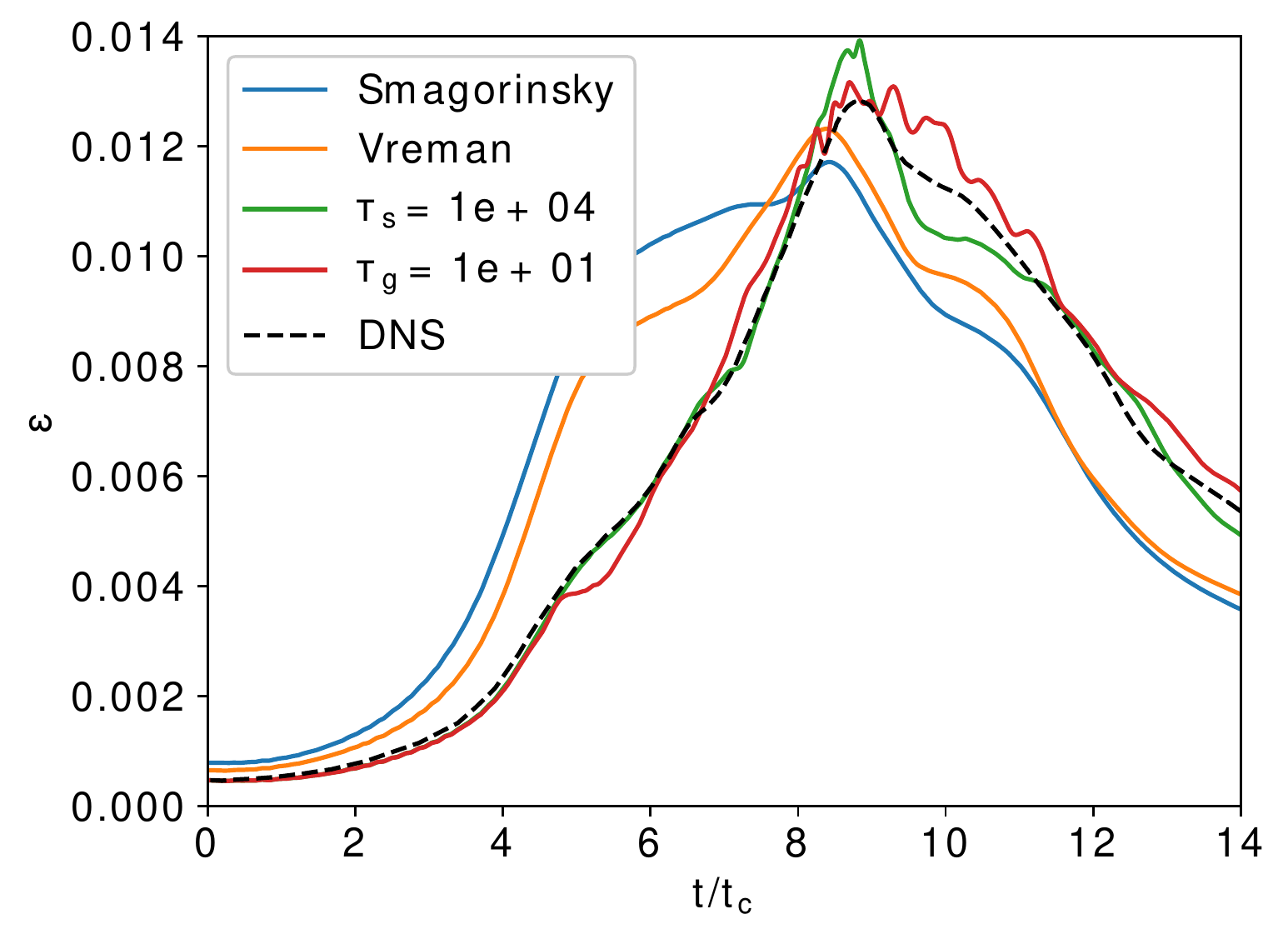}
		\caption{}
	\end{subfigure}
	\begin{subfigure}{.45\textwidth}
		\includegraphics[width=180pt]{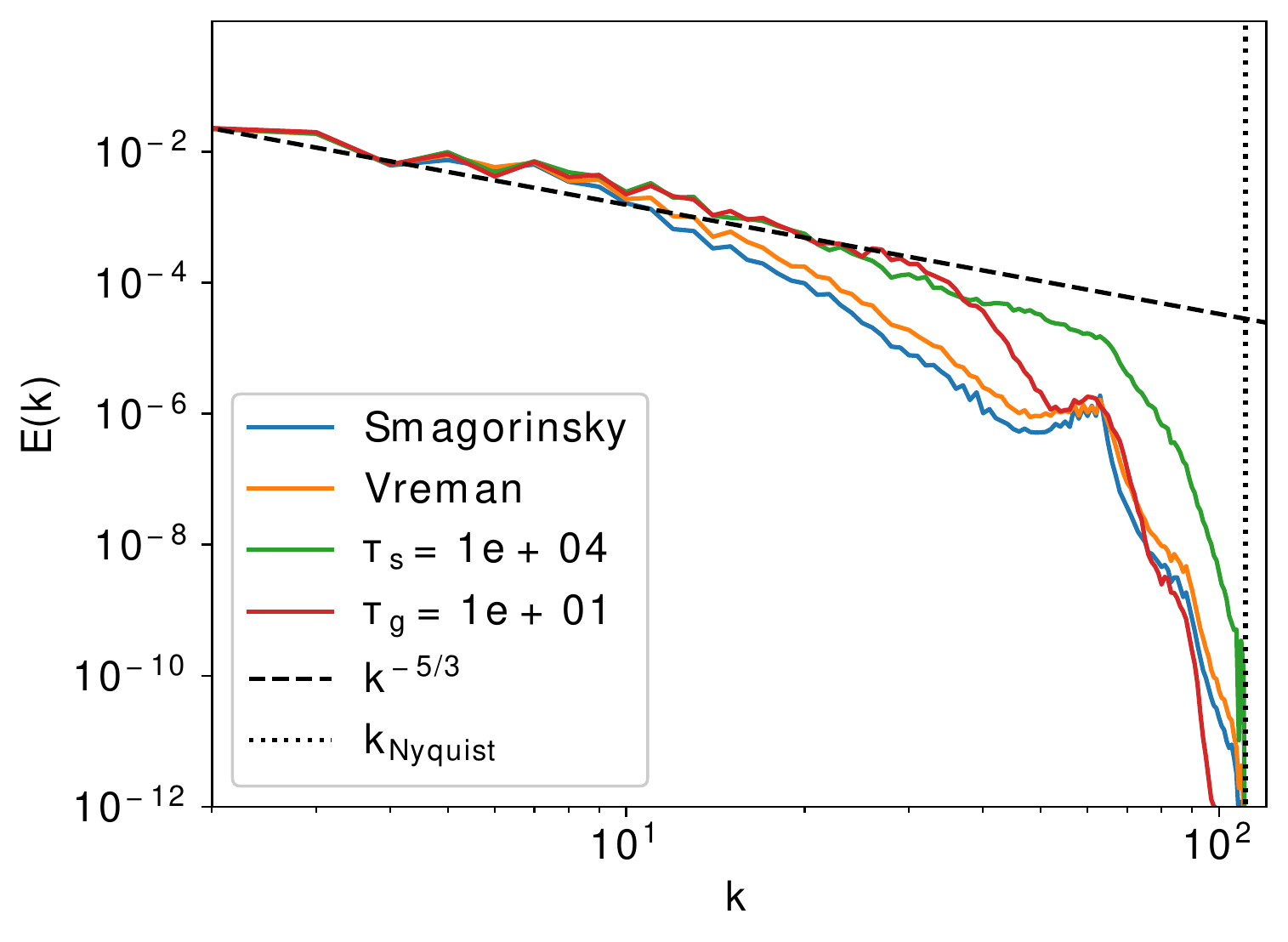}
		\caption{}
	\end{subfigure}
	\centering
	\caption{Taylor Green vortex: Comparison with of jump penalty strategy with traditional LES SGS model. a) Energy dissipation rate. b) Energy spectra at $t/t_c=8$.}
	\label{fig:TGV-dissipation3}
\end{figure*}

Finally, we compare both jump penalty strategies (optimal penalty parameters) with the traditional LES Sub-Grid Scale (SGS) model (Smagorinsky \cite{Smagorinsky} or Vreman \cite{vreman} adapted to DG) in figure \ref{fig:TGV-dissipation3}. Both the energy dissipation rate with respect to time (Figure \ref{fig:TGV-dissipation3}a) and the energy spectra (Figure \ref{fig:TGV-dissipation3}b) show the clear superiority of the jump penalty strategies over classic LES closure models. Figure \ref{fig:TGV-dissipation3}a shows that classic closure models provide a lower maximum peak and a higher energy dissipation rate at earlier times than the peak $t/t_c<8$, compared to DNS results.
Figure \ref{fig:TGV-dissipation3}b shows that classic closure models dissipate significantly at midwave numbers, while jump penalties maintain the correct energy dissipation for higher wavenumbers. In addition, we observe that the solution jump allows higher wavenumbers to be captured. We conclude that both gradient and solution jump penalisation stabilise under-resolved simulations and improve the simulations, when compared to the original unpenalised scheme and to classic explicit subgrid models (Smagorisnky and Vreman).

\section{Conclusions}
\label{sec:conclusion}
We have studied and compared two newly proposed techniques to stabilise underresolved LES in high-order DG schemes. We analyse the numerical characteristics for two variants of the jump penalty stabilisation, one penalising the solution (proposed in \cite{ferrer2017interior}) and another penalising the solution gradient (proposed in \cite{moura2022gradient,MOURA2022114200}). We analyse these schemes using a linear eigensolution analysis, a 1D nonlinear Burgers equation (mimicking a turbulent cascade), and 3D turbulent Navier-Stokes through the Taylor-Green Vortex problem. Main conclusions are drawn as follows:

\begin{itemize}

  \item \textbf{Eigensolution / von Neumann analysis}: We show that both penalty terms enhance dissipation at high wavenumbers and improve dispersion behaviours at mid- to high wavenumbers. When optimal penalties are used, the addition of the penalty significantly improves the baseline scheme (no penalty).

  \item \textbf{1D Burgers turbulent problem}: The results of von Neumann analysis are consistent with this non-linear test case. Both penalty terms improve the baseline solution (no penalty). Within a reasonable range of penalty terms, the energy decay of the unforced Burgers turbulence is well described, and the resolved wavenumber range is enlarged. When penalising the solution, the results are less sensitive to the chosen constant or penalty term accompanying the solution jump. When penalising the gradients, we observe a larger variation in the results depending on the penalty parameter and the polynomial order. While both jump penalties improve the baseline solution, the optimal gradient penalty seems to provide slightly better resolution at the highest wavenumbers. 

  \item \textbf{3D turbulent Taylor-Green Vortex}: This turbulent 3D case consolidates the previous findings. Both jump penalisations improve the baseline scheme. In addition, we also show that they outperform classic LES models (Smagorisky and Vreman). The optimal penalty for the solution and gradient jumps are the highest tested. This facilitates the estimation of this parameter but can lead to CFL-type restrictions. In the future, this term will be treated implicitly to limit its stiffness for large penalties.  
\end{itemize}

Future work will test these schemes for wall bounded turbulence and complex geometries.

\section*{Acknowledgements}
Esteban Ferrer would like to thank the support of the Spanish Minister MCIN/AEI/10.13039/501100011033 and the European Union NextGenerationEU/PRTR for the grant "Europa Investigación 2020" EIN2020-112255, and also the Comunidad de Madrid through the call Research Grants for Young Investigators from the Universidad Politécnica de Madrid. Additionally, the authors acknowledge the Universidad Politécnica de Madrid (www.upm.es) for providing computing resources on the Magerit Supercomputer. 
\color{black}

\appendix
\section{Eigensolution analysis}
\label{sec:appendEA}
To apply eigensolution analyses, we first obtain the analytical solution for the advection problem, initialised by a harmonic wave $u(x,0) = \text{exp}(ikx)$ under the periodic boundary conditions:

\begin{equation}
    u(x,t) = \text{exp} [i(kx -\omega t)],
\end{equation}
where $k$ is the wave frequency, $\omega = a k$ is the (angular) frequency, and $i = \sqrt{-1}$ is the imaginary unit. Substituting the analytical solution into the advection, we obtain the classical dispersion relationship:

\begin{equation}
    k = \tilde{\omega} = \omega / a.
\end{equation}

We can perform the temporal eigensolution analysis considering $k$ as a real value and $\tilde{\omega} = \tilde{\omega}(k)$, and the spatial eigensolution analysis by considering $\tilde{\omega}$ as a real value and $ k = k (\tilde{\omega})$.

\subsection{Theory}
For temporal eigensolution analysis, periodic boundary conditions are considered. Therefore, for each element with index $n$, we have $\hat{\mathbf{u}}_{n-1} = e^{-ikh}\hat{\mathbf{u}}_{n}$ and $\hat{\mathbf{u}}_{n+1} = e^{+ikh}\hat{\mathbf{u}}_{n}$. The DG space discretization for this element results in the following semi-discrete formulation:

\begin{equation}
\label{eq:L-C-R}
    \frac{d\hat{\mathbf{u}}_n}{dt} = \left ( \mathbf{L} e^{-ikh} +  \mathbf{C}  +  \mathbf{R} e^{+ikh} \right ) \hat{\mathbf{u}}_{n} = \mathbf{M} \hat{\mathbf{u}}_{n},
\end{equation}
where $\mathbf{L}$, $\mathbf{C}$ and $\mathbf{R}$ refer to the operator matrices from the left, central, and the right elements, respectively. The specific formulation and details of these matrices are given in \ref{sec:appendDG}. This semi-discrete matrix formulation is the basis for the eigensolution analysis. It can be transformed into the compact form of an eigenvalue problem:

\begin{equation}
    - i \omega^* \hat{\mathbf{u}}_n = \mathbf{M} \hat{\mathbf{u}}_{n},
\end{equation}
where $\omega^* = a k^*$ becomes a complex value due to the dispersion and dissipation errors of the space discretisation. This wavenumber $\omega^*$ relates to the eigendecomposition of the coefficient matrix $\mathbf{M}$ (with $P+1$ solutions):

\begin{equation}
\label{eq:LCR}
    - i \omega^*_m = \lambda_m \\, \ \lambda_m \mathbf{v}_m = \mathbf{M} \mathbf{v}_m,
\end{equation}
where $\lambda_m$ and $\mathbf{v}_m$ are the $m$th eigenvalues and eigenvectors of matrix $\mathbf{M}$, respectively. By defining the modified wavenumber $k^*_m$ for each $\omega^*_m$ and $\lambda_m$, we have:

\begin{equation}
    \text{Real}(k^*_m) = \frac{\text{Real}(\omega^*_m)}{a} = -\frac{\text{Imag}(\lambda_m)}{a}\\, \ \text{Imag}(k^*_m) = \frac{\text{Imag}(\omega^*_m)}{a} = \frac{\text{Real}(\lambda_m)}{a}.
\end{equation}

For the advection equation, the difference between $\text{Real}(k^*)$ and $k$ is due to the dispersion error, indicating the change in the wavenumber of the solution. The difference between $\text{Imag}(k^*)$ and $0$ corresponds to the dissipation (diffusion) error, where $\text{Imag}(k^*) \leq 0$ holds for a stable scheme. Note that for the analytical solution we have $\text{Real}(k^*) = k$ and $\text{Imag}(k^*) = 0$. In addition, $\mathbf{M}$ will have $P+1$ eigenmodes, where the one that recovers $k^* = k$ as $k \rightarrow 0$ will be identified as the physical mode, while others are secondary modes, which are in fact translated replicas of the primary mode whose properties are essentially contained in the physical (primary) mode \citep{moura2015linear}.

\subsection{Discontinuous Galerkin for advection equation in matrix form}
\label{sec:appendDG}
For eigensolution analysis, considering the limit of a very large Péclet number, the following advection equation is studied:

\begin{equation}
    \frac{\partial u}{\partial t} + \frac{\partial f}{\partial x} = 0,
\end{equation}
where $f=au$ is the advection flux. Following Section \ref{sec:numerical}, the computational domain is discretised into nonoverlapping elements, and the solution is represented by a modal formulation in Eq. \ref{eq:poly}. By projecting the equation onto the test basis $\phi_i$, we have the following

\begin{equation}
    \int_{\Omega_n}^{} \phi_i \left (\frac{\partial u}{\partial t} +  \frac{\partial f}{\partial x} \right ) dx = 0.
\end{equation}
Through integration by part and using the orthogonality of the basis functions, one can arrive at

\begin{equation}
\label{eq:app-1}
    \frac{h}{2} \frac{\partial \hat{u}_i}{\partial t} = \int_{\Omega_n}^{}  \frac{\partial \phi_i}{\partial \xi} f d\xi  - (f^* \phi_i)|_{\Omega_n^R} + (f^* \phi_i)|_{\Omega_n^L} ,
\end{equation}
where $\Omega_n^L$ and $\Omega_n^R$ denote the left and right interfaces of the current element. The numerical flux is a function of the values from both sides of the interface (left $u^-$ and right $u^+$), which takes the following form:

\begin{equation}
    f^{*}(u^{-}_,u^{+}) = \frac{f(u^{-})+f(u^{+})}{2} + \frac{\lambda}{2} \left| \frac{\partial f}{\partial u} \right | (u^{-}-u^{+}),
\end{equation}
where $\frac{\partial f}{\partial u} = a$. $\lambda$ is the upwinding parameter, where $\lambda = 0$ and $\lambda = 1$ refer to the central and upwind flux, respectively. Note that for the left interface $\Omega_n^L$, $u^{-}$ and $f(u^{-})$ come from the left element, while for the right element $\Omega_n^R$, $u^{+}$ and $f(u^{+})$ come from the right element. The numerical fluxes on both interfaces of the element can be expressed as:

\begin{equation}
    f^{*}|_{\Omega_n^L} = \frac{1}{2}(a + \lambda |a|) \mathbf{\phi}^T(1) \hat{\mathbf{u}}_{n-1}
   + \frac{1}{2}(a - \lambda |a|) \mathbf{\phi}^T(-1) \hat{\mathbf{u}}_{n} ,
\end{equation}

\begin{equation}
    f^{*}|_{\Omega_n^R} = \frac{1}{2}(a + \lambda |a|) \mathbf{\phi}^T(1) \hat{\mathbf{u}}_{n}
   + \frac{1}{2}(a - \lambda |a|) \mathbf{\phi}^T(-1) \hat{\mathbf{u}}_{n+1},
\end{equation}
where $\mathbf{\phi}$ refers to a vector of modal basis functions, and $\hat{\mathbf{u}}$ refers to a vector of modal coefficients of each element. Combining the above equations, Equation \ref{eq:app-1} can be written as follows:

\begin{equation}
\begin{aligned}
    \frac{h}{2} \frac{\partial \hat{\mathbf{u}}_n}{\partial t} = & a \mathbf{S} \hat{\mathbf{u}}_n + \\
    & \mathbf{\phi}(-1) \left (\frac{1}{2}(a + \lambda |a|) \mathbf{\phi}^T(1) \hat{\mathbf{u}}_{n-1}
   + \frac{1}{2}(a - \lambda |a|) \mathbf{\phi}^T(-1) \hat{\mathbf{u}}_{n} \right ) - \\
   & \mathbf{\phi}(+1) \left( \frac{1}{2}(a + \lambda |a|) \mathbf{\phi}^T(1) \hat{\mathbf{u}}_{n}
   + \frac{1}{2}(a - \lambda |a|) \mathbf{\phi}^T(-1) \hat{\mathbf{u}}_{n+1} \right),
\end{aligned}
\end{equation}

Referring to the semidiscrete formulation in Equation \ref{eq:L-C-R}, the left, middle, and right operators $\mathbf{L}$, $\mathbf{C}$, $\mathbf{R}$ are written as:

\begin{equation}
\begin{aligned}
    \mathbf{L} = \frac{1}{h} \mathbf{\phi}(-1) \left [(a + \lambda |a|) \mathbf{\phi}^T(1) \right ],
\end{aligned}
\end{equation}

\begin{equation}
\begin{aligned}
    \mathbf{C} = \frac{2}{h} \left[ a \mathbf{S} + \mathbf{\phi}(-1) \left( \frac{1}{2}(a - \lambda |a|) \mathbf{\phi}^T(-1) \right) - \mathbf{\phi}(+1) \left( \frac{1}{2}(a + \lambda |a|) \mathbf{\phi}^T(1) \right) \right],
\end{aligned}
\end{equation}

\begin{equation}
\begin{aligned}
    \mathbf{R} = - \frac{1}{h} \mathbf{\phi}(+1) \left[ \frac{1}{2}(a - \lambda |a|) \mathbf{\phi}^T(-1) \right].
\end{aligned}
\end{equation}

Furthermore, from Equations \ref{eq:gradJP} and \ref{eq:SolJP}, additional operators are derived for the the jump penalty stabilisation. Gradient jump penalty leads to the following matrices:

\begin{equation}
\begin{aligned}
    \mathbf{L}_{gradient} = \frac{4 \tau_g}{h^2} \mathbf{\phi}_{\xi}(-1) \mathbf{\phi}_{\xi}^T(1) ,
\end{aligned}
\end{equation}

\begin{equation}
\begin{aligned}
    \mathbf{C}_{gradient} = - \frac{4 \tau_g}{h^2} \left[ \mathbf{\phi}_{\xi}(-1) \mathbf{\phi}_{\xi}^T(-1) + \mathbf{\phi}_{\xi}(+1) \mathbf{\phi}_{\xi}^T(+1) \right ],
\end{aligned}
\end{equation}

\begin{equation}
\begin{aligned}
    \mathbf{R}_{gradient} = \frac{4 \tau_g}{h^2}  \mathbf{\phi}_{\xi}(+1) \mathbf{\phi}_{\xi}^T(-1),
\end{aligned}
\end{equation}
where $\mathbf{\phi}_{\xi} = \frac{\partial \mathbf{\phi}}{\partial \xi}$ is the gradient of the modal basis function. The solution jump penalty leads to the following matrices:

\begin{equation}
\begin{aligned}
    \mathbf{L}_{solution} = \frac{2 \tau_s }{h} \mathbf{\phi}(-1) \mathbf{\phi}^T(1) ,
\end{aligned}
\end{equation}

\begin{equation}
\begin{aligned}
    \mathbf{C}_{solution} = - \frac{2 \tau_s }{h} \left[ \mathbf{\phi}(-1) \mathbf{\phi}^T(-1) + \mathbf{\phi}(+1) \mathbf{\phi}^T(+1) \right ],
\end{aligned}
\end{equation}

\begin{equation}
\begin{aligned}
    \mathbf{R}_{solution} = \frac{2 \tau_s}{h}  \mathbf{\phi}(+1) \mathbf{\phi}^T(-1),
\end{aligned}
\end{equation}

\section{Additional results for eigensolution analyses}
\label{sec:app-P2}
\subsection{Central Riemann flux}
\begin{figure*}[htbp]
\begin{subfigure}{.45\textwidth}
		\includegraphics[width=180pt]{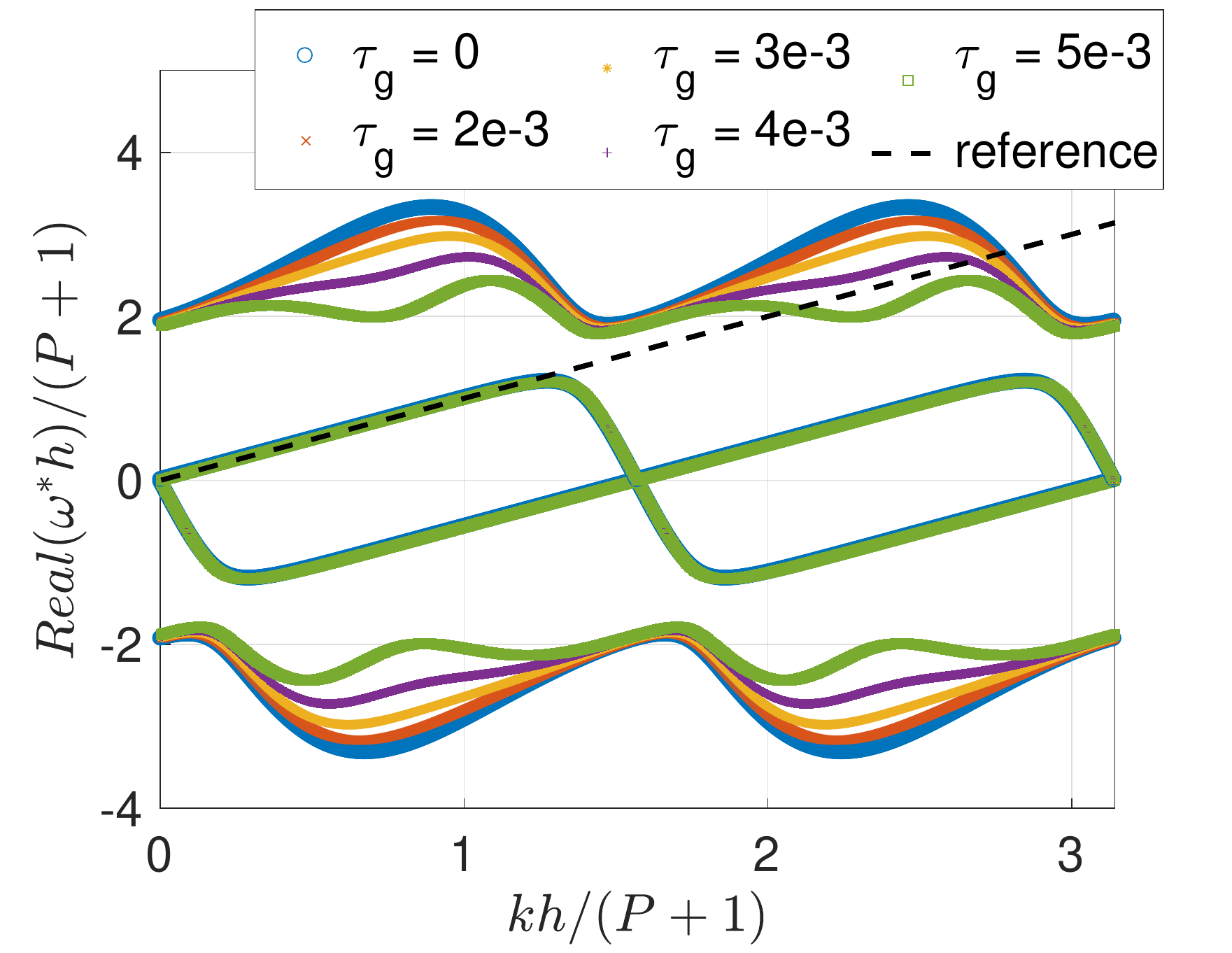}
		\caption{}
	\end{subfigure}
	\begin{subfigure}{.45\textwidth}
		\includegraphics[width=180pt]{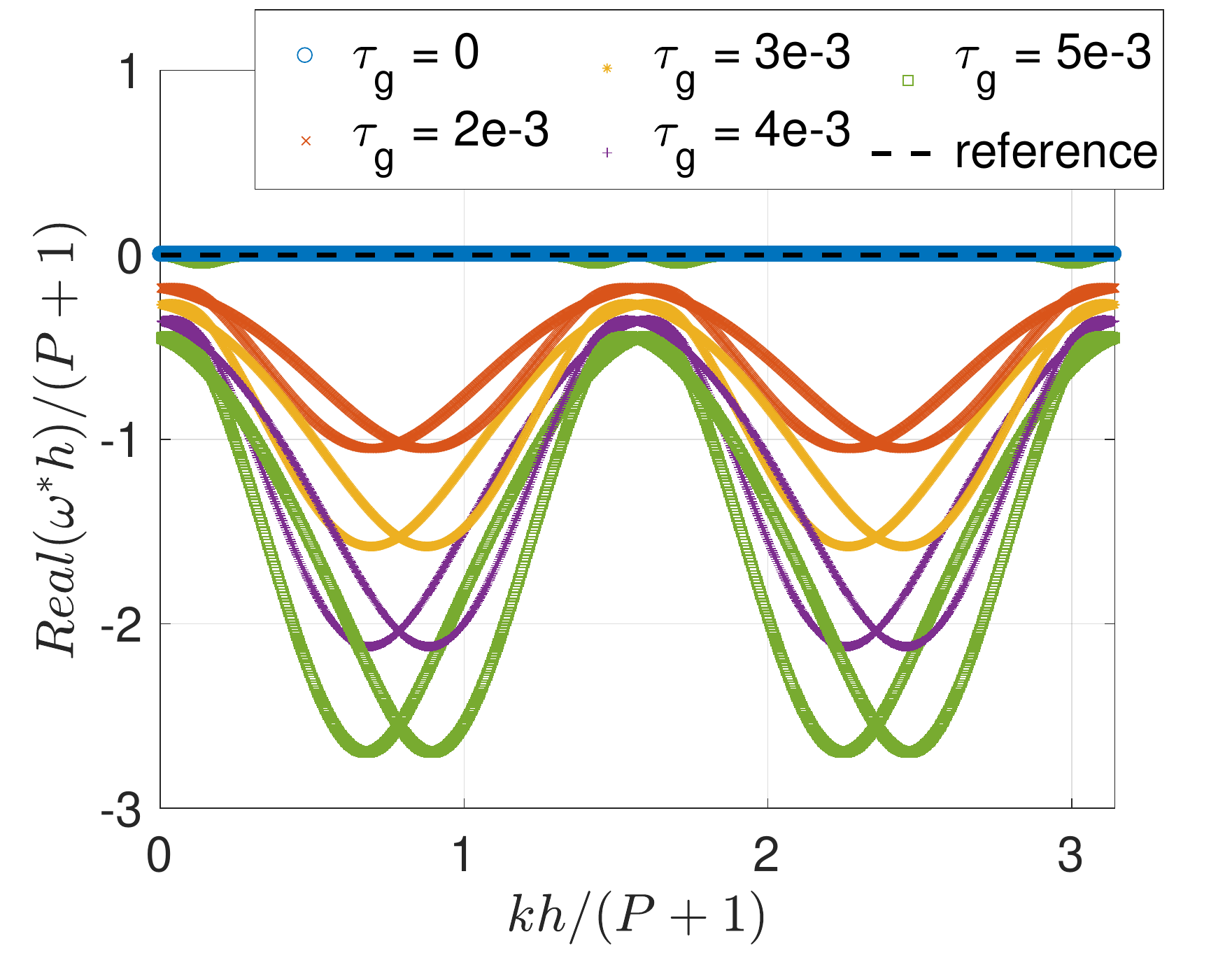}
		\caption{}
	\end{subfigure}
	\centering
	\caption{Dispersion-dissipation behaviour with $P = 3$ and central flux (gradient jump penalty). a) Dispersion. b) Dissipation.}
	\label{fig:P3-grad-cen}
\end{figure*}

\begin{figure*}[htbp]
\begin{subfigure}{.45\textwidth}
		\includegraphics[width=180pt]{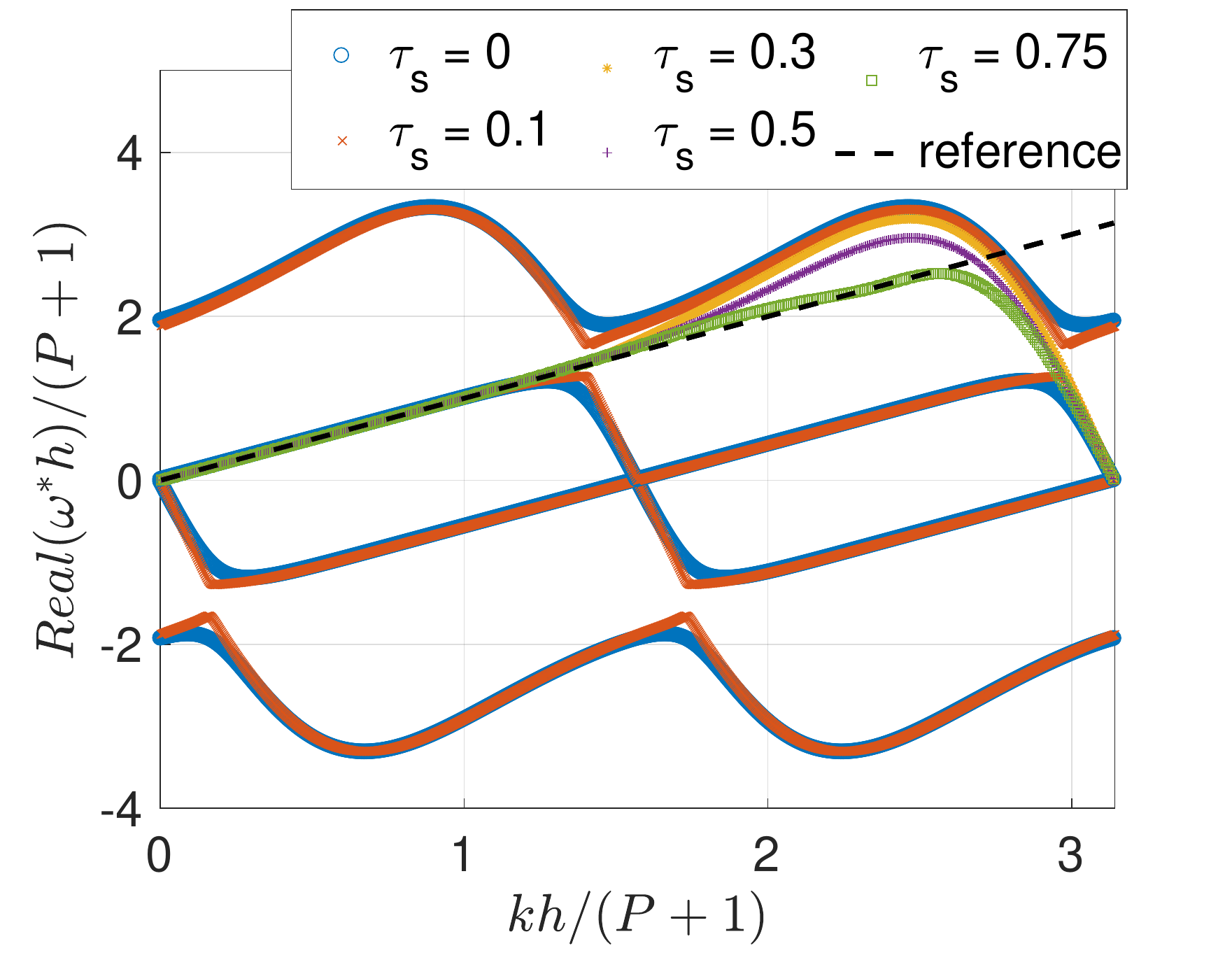}
		\caption{}
	\end{subfigure}
	\begin{subfigure}{.45\textwidth}
		\includegraphics[width=180pt]{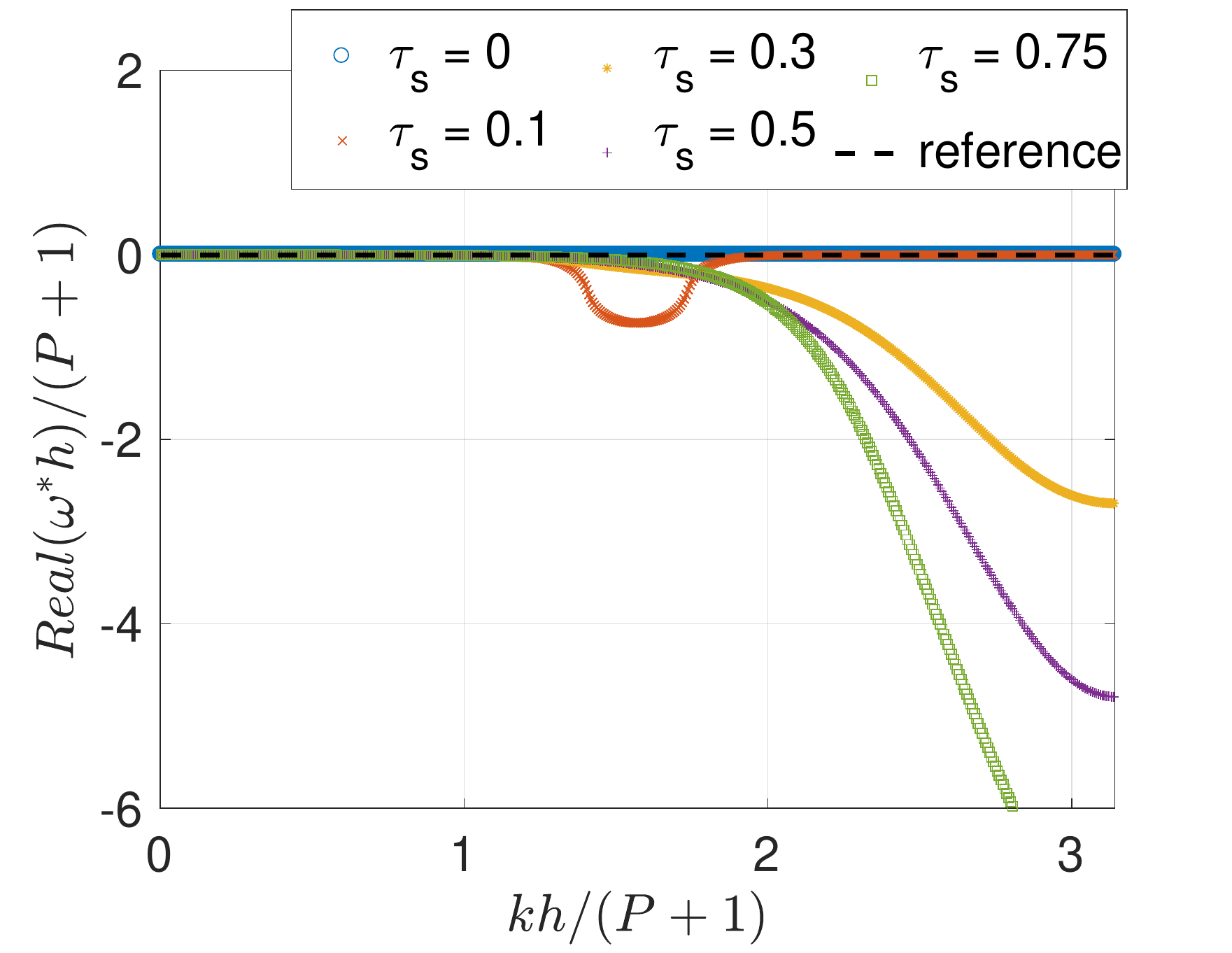}
		\caption{}
	\end{subfigure}
	\centering
	\caption{Dispersion-dissipation behaviour with $P = 3$ and central flux (solution jump penalty). a) Dispersion. b) Dissipation (physical mode).}
	\label{fig:P3-sol-cen}
\end{figure*}

In this subsection, the influence of penalty parameters $\tau_g$ and $\tau_s$ is studied for polynomial order P = 3 and a central Riemann flux. Results are shown in Figure \ref{fig:P3-grad-cen} ($P=3$, gradient jump penalty) and Figure \ref{fig:P3-sol-cen} ($P=3$, solution jump penalty). As discussed in \cite{hu1999analysis,van2007dispersion,asthana2015high}, when central flux is used, standard DG shows zero dissipation for all numerical modes (including both physical and spurious modes). For standard DG, the only source of dissipation comes from the upwind flux in the Riemann solver used at the element interface \cite{asthana2015high}. In terms of dispersion behaviour, the physical mode (which follows the reference dispersion curve at small wavenumbers $kh/(P+1) < 1.1$) exhibits a negative wave speed in the resolved wavenumber range $kh/(P+1) > 1.57$. However, as analysed by Asthana and Jameson \cite{asthana2015high}, the energy of the physical mode falls rapidly when the dispersion curve is not well traced $k/(P+1) > 1.1$, while one of the spurious modes starts to dominate the energy beyond this wavenumber (the upper blue curve in Figure \ref{fig:P3-grad-cen}a). This spurious mode has small dispersion errors afterward; therefore, the overall performance of central flux is maintained by this mode. 

In Figure \ref{fig:P3-grad-cen}a, when the gradient jump penalty is introduced, the physical mode is not influenced, while the spurious mode dominating the numerical behaviour at large wavenumbers is modified. As $\tau_g$ increases, a spurious mode with better dispersion property at high wavenumber can be obtained (e.g., $\tau_g = 4 \times 10^{-3}$). However, an optimal $\tau_g$ exists and the dispersion property becomes worse beyond this value. It can also be observed that an additional dissipation effect is induced for gradient jump stabilisation, thus stabilising the DG simulation with a central scheme. However, there is still a gap in dispersion near the medium wavenumber range, where the dominant mode switches from the physical mode to one spurious mode. The effect of the solution jump penalty stabilisation is shown in Figure \ref{fig:P3-sol-cen}. As $\tau_s$ increases, the physical mode will dominate across all wavenumbers, therefore only the physical mode is depicted for the dissipation curve. Dissipation is introduced and increases as the wavenumber increases, resulting in the same trend as using upwind fluxes. In fact, it is found that when $\tau_s = 0.5$, dispersion-dissipation behaviour is identical to that of standard DG with upward Riemann fluxes ($\lambda = 1$, defined in \ref{sec:appendDG}).  This indicates that solution jump penalisation compensates the upwinding effect of the Riemann solver. In particular, when $\tau_s = 0.75$, the physical mode shows a beneficial dispersion property which follows the right wavenumber at a very high wavenumber (around $kh/(P+1) = 2.5$). The dissipation behaviour is also improved, where small dissipation is maintained from low to medium wavenumber, and large dissipation is shown at large wavenumbers. Generally, both results show that jump penalty stabilisation improves the dispersion and dissipation behaviours when a central Riemann flux is used. As a complementary reference, the optimal penalty parameters for the central fluxes are summarised in Table \ref{table1}. Compared to upwind fluxes, the optimal $\tau_s$ is increased by $0.5$, due to the lack of upwinding in the Riemann solver. Optimal $\tau_g$ becomes positive to offer upwind effects to stabilise the numerical schemes.

\begin{table}[htbp!]
	\vspace{20pt}
	\centering
	\begin{tabular}{p{2cm}p{2cm}p{2cm}p{2cm}p{2cm}p{2cm}p{2cm}}
  \hline
 Polynomial & \multicolumn{1}{c}{P = 2} & \multicolumn{1}{c}{P = 3} & \multicolumn{1}{c}{P = 4}\\
  \hline
  Gradient & $1.2 \times 10^{-2}$ & $4 \times 10^{-3}$ & $2 \times 10^{-3}$\\
  Solution & $0.75$ & $0.75$ & $0.75$\\
  \hline
 \end{tabular}
	\caption{Optimal penalty parameters of advection equation for different polynomial orders (central Riemann flux).}
	\label{table1}
\end{table}

\subsection{Other polynomial orders}
\begin{figure*}[htbp]
\begin{subfigure}{.45\textwidth}
		\includegraphics[width=180pt]{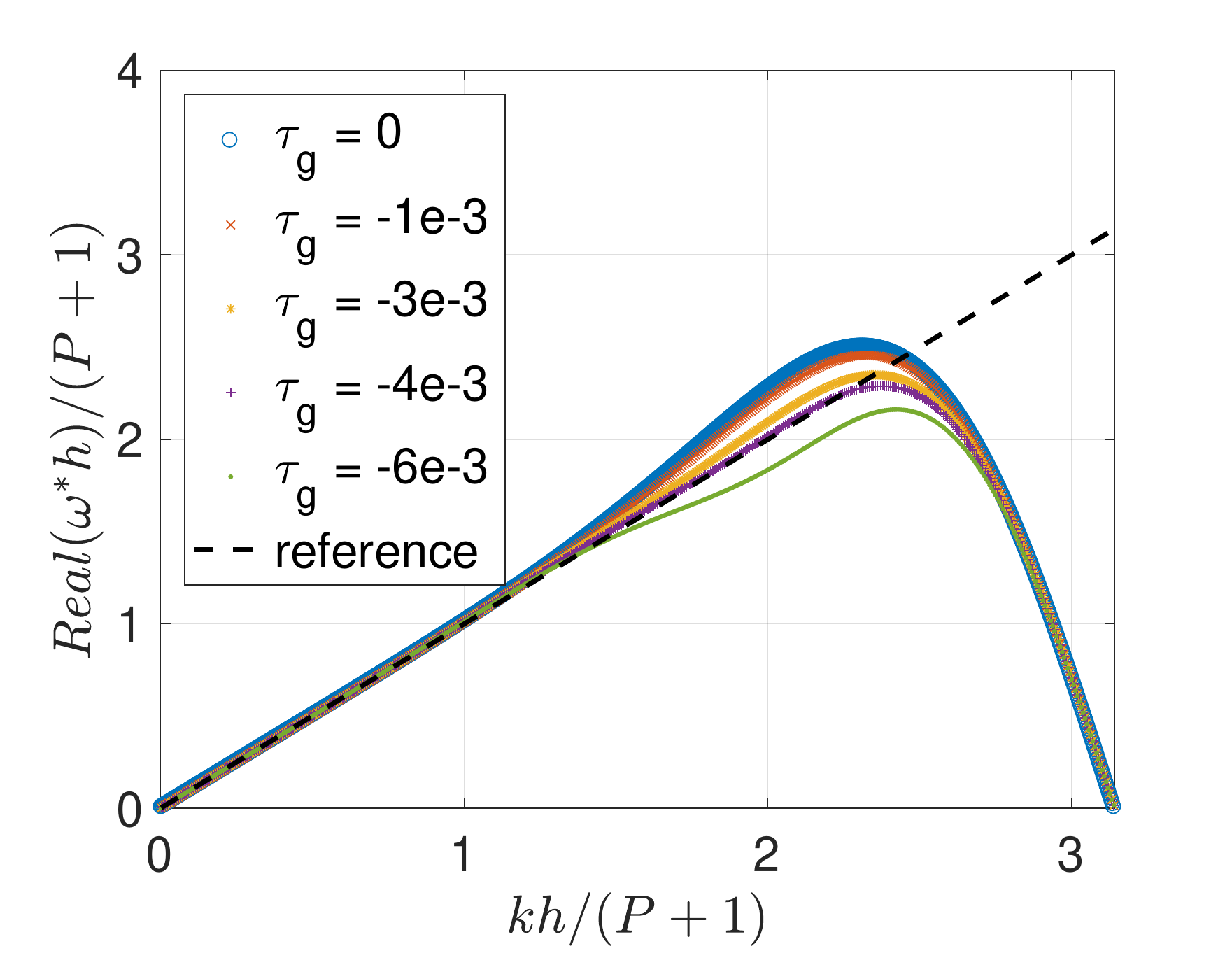}
		\caption{}
	\end{subfigure}
	\begin{subfigure}{.45\textwidth}
		\includegraphics[width=180pt]{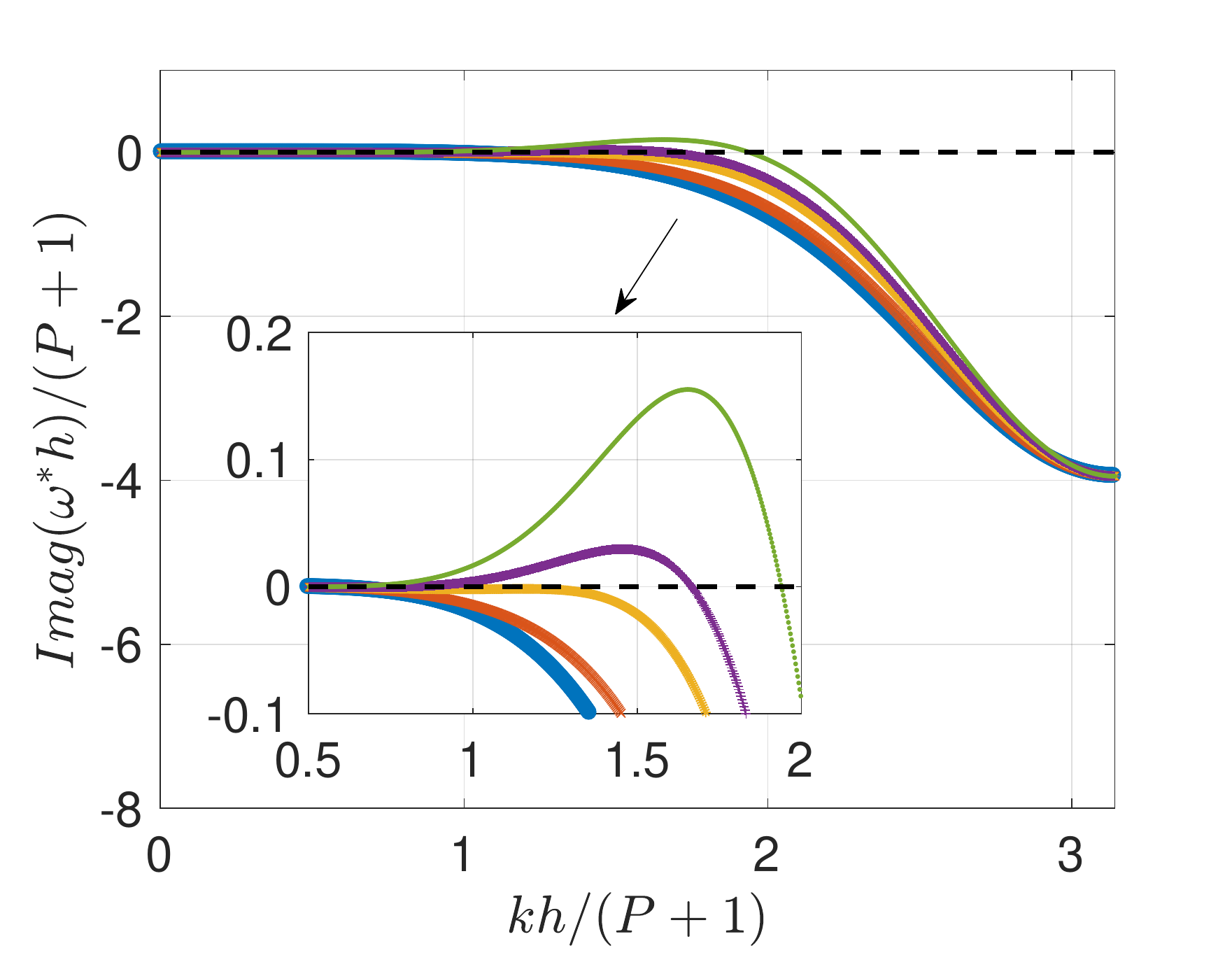}
		\caption{}
	\end{subfigure}
	\centering
	\caption{Dispersion-dissipation behaviour with $P = 2$ and upwind flux (gradient jump penalty). a) Dispersion. b) Dissipation.}
	\label{fig:P2-grad}
\end{figure*}

\begin{figure*}[htbp]
\begin{subfigure}{.45\textwidth}
		\includegraphics[width=180pt]{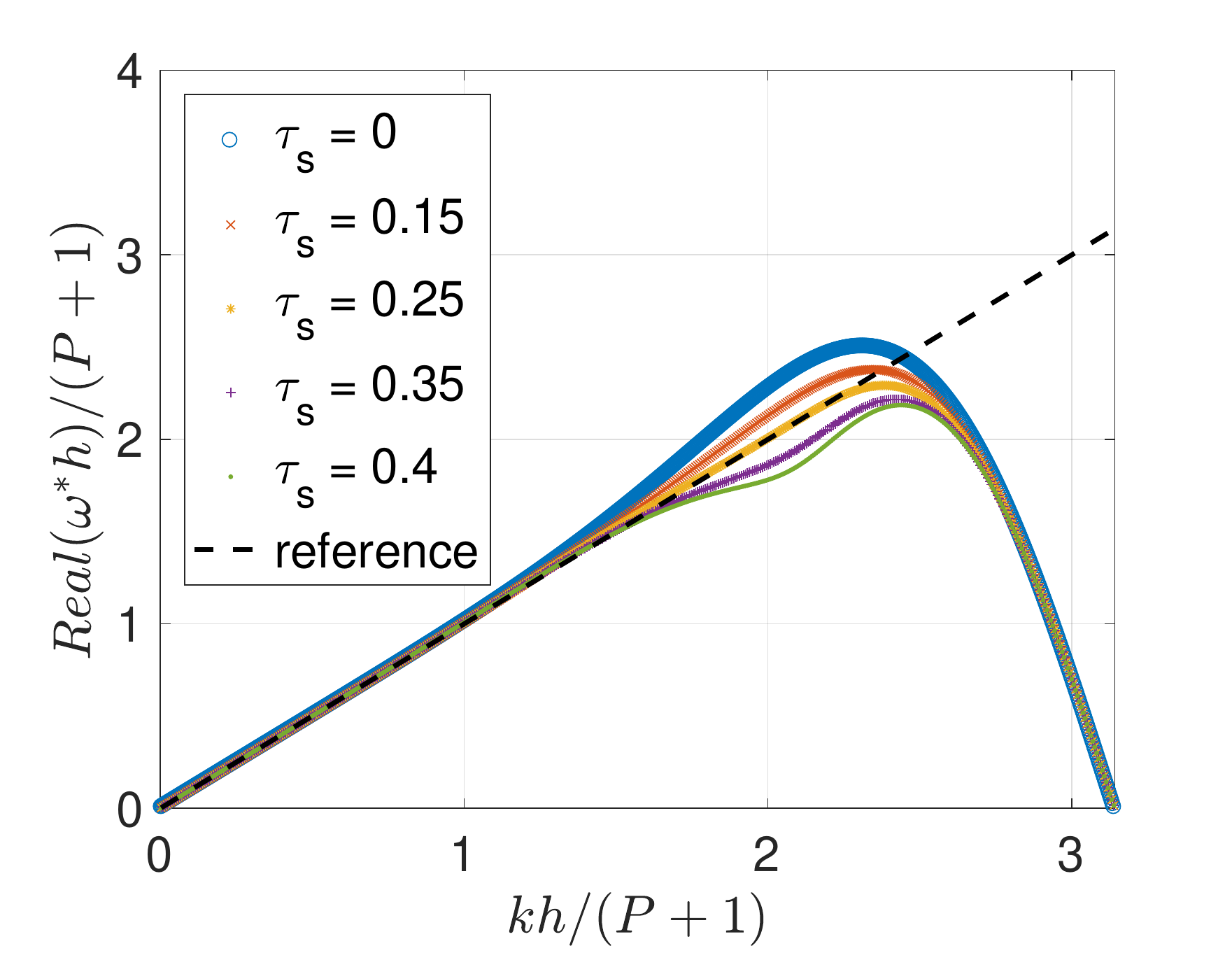}
		\caption{}
	\end{subfigure}
	\begin{subfigure}{.45\textwidth}
		\includegraphics[width=180pt]{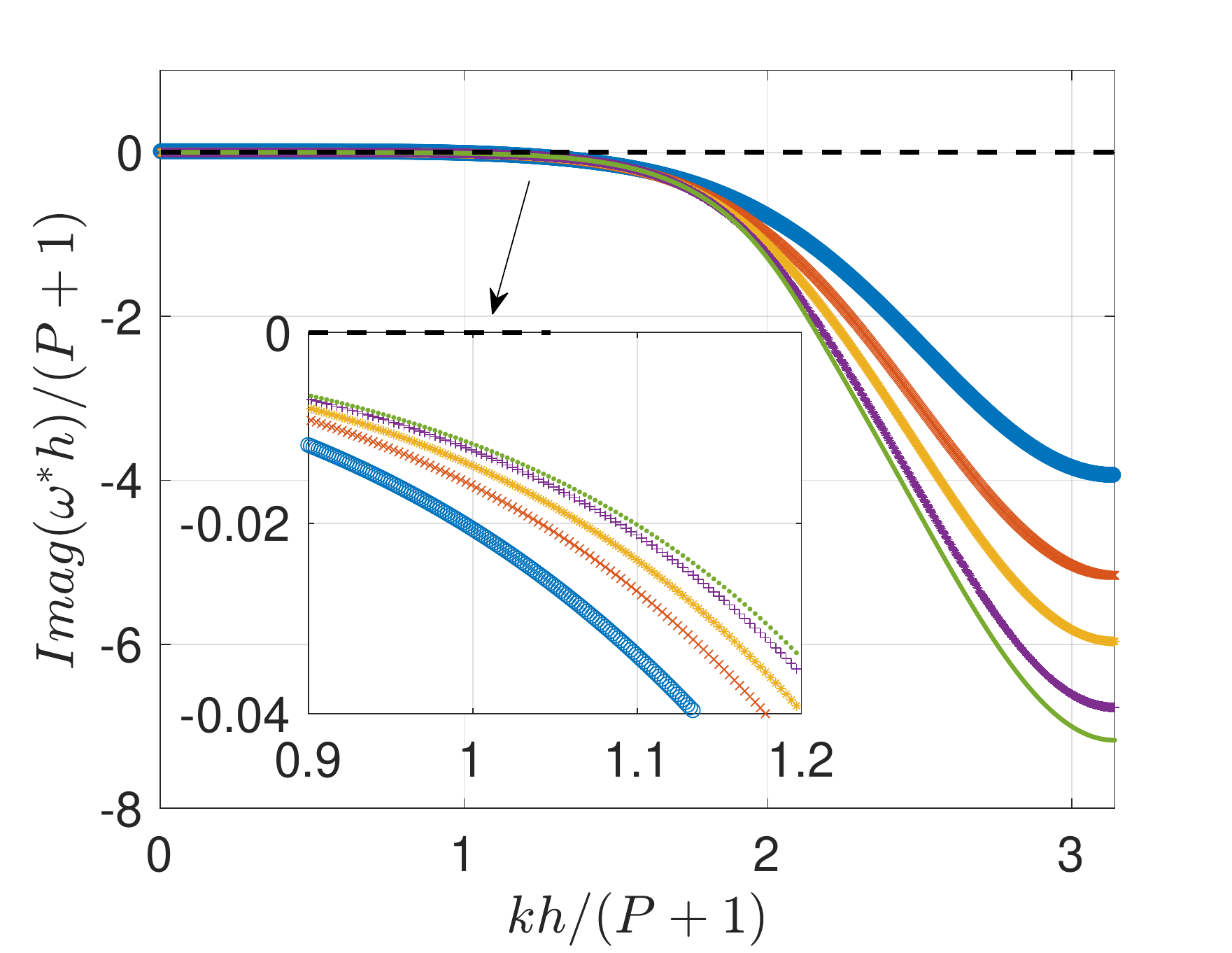}
		\caption{}
	\end{subfigure}
	\centering
	\caption{Dispersion-dissipation behaviour with $P = 2$ and upwind flux (solution jump penalty). a) Dispersion. b) Dissipation.}
	\label{fig:P2-sol}
\end{figure*}

This section gives additional results on eigensolution analysis of jump penalty stabilisation methods for polynomial orders $P=2$ and $P=4$. Similar conclusions can be drawn compared to the results for $P=3$. For upwind fluxes, both methods of jump penalty stabilisation improve the standard DG scheme with a proper selection of penalty parameter. Typical results for $P=2$ are shown in Figures \ref{fig:P2-grad} and \ref{fig:P2-sol}. It should be noted that optimal jump penalty terms improve the dispersion behaviour at medium- to high-wavenumbers. However, as shown in \ref{fig:P2-grad}b, for the gradient jump penalty stabilisation, instability is induced when the optimal dispersion behaviour is reached, which limits the penalty parameter $\tau_g$. For the solution jump penalty stabilisation, optimal dispersion property is reached when $\tau_s = 0.25$, where a higher frequency can be resolved that improves the dispersion property. The dissipation property continues to improve as $\tau_s$ increases, where dissipation is small for a medium wavenumber, while large dissipation is offered for a high wavenumber to mimic the physical dissipation effects. 

\begin{figure*}[htbp]
\begin{subfigure}{.45\textwidth}
		\includegraphics[width=180pt]{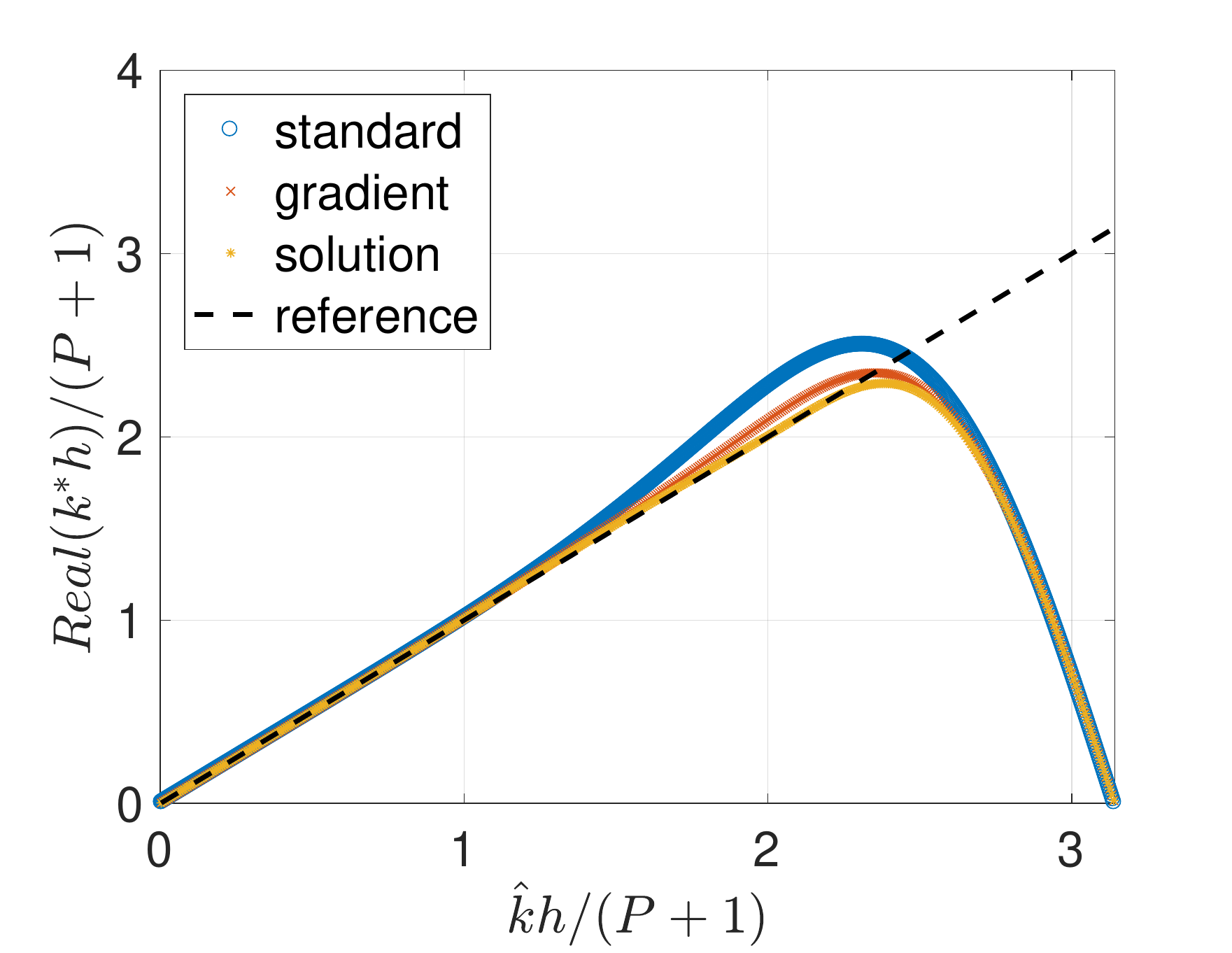}
		\caption{}
	\end{subfigure}
	\begin{subfigure}{.45\textwidth}
		\includegraphics[width=180pt]{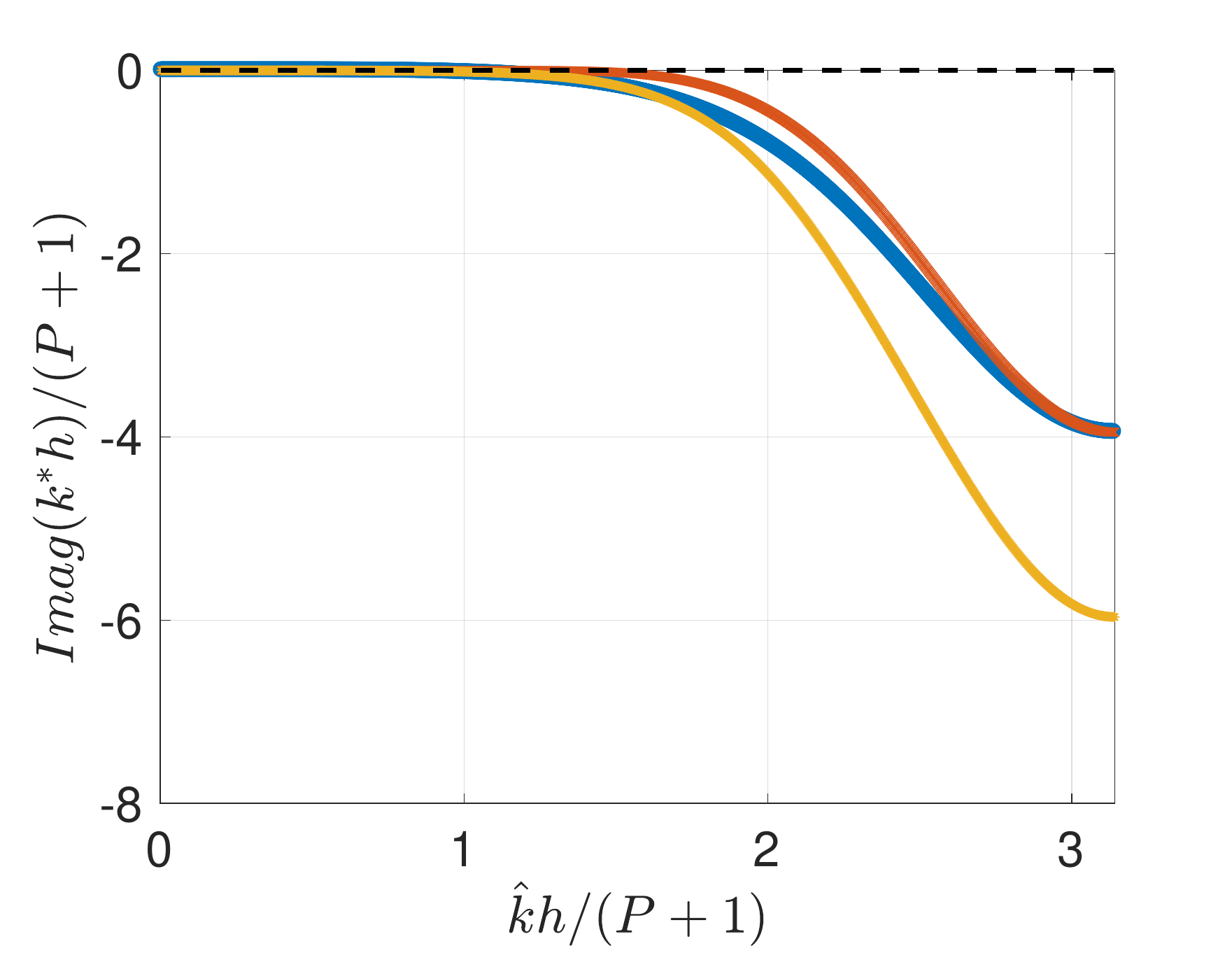}
		\caption{}
	\end{subfigure}
	\centering
	\caption{Optimal dispersion-dissipation behaviour at P = 2 (upwind flux, $\tau_g = -3\times 10^{-3}, \tau_s = 0.25$). a) Dispersion. b) Dissipation.}
	\label{fig:P2-comp}
\end{figure*}

\begin{figure*}[htbp]
\begin{subfigure}{.45\textwidth}
		\includegraphics[width=180pt]{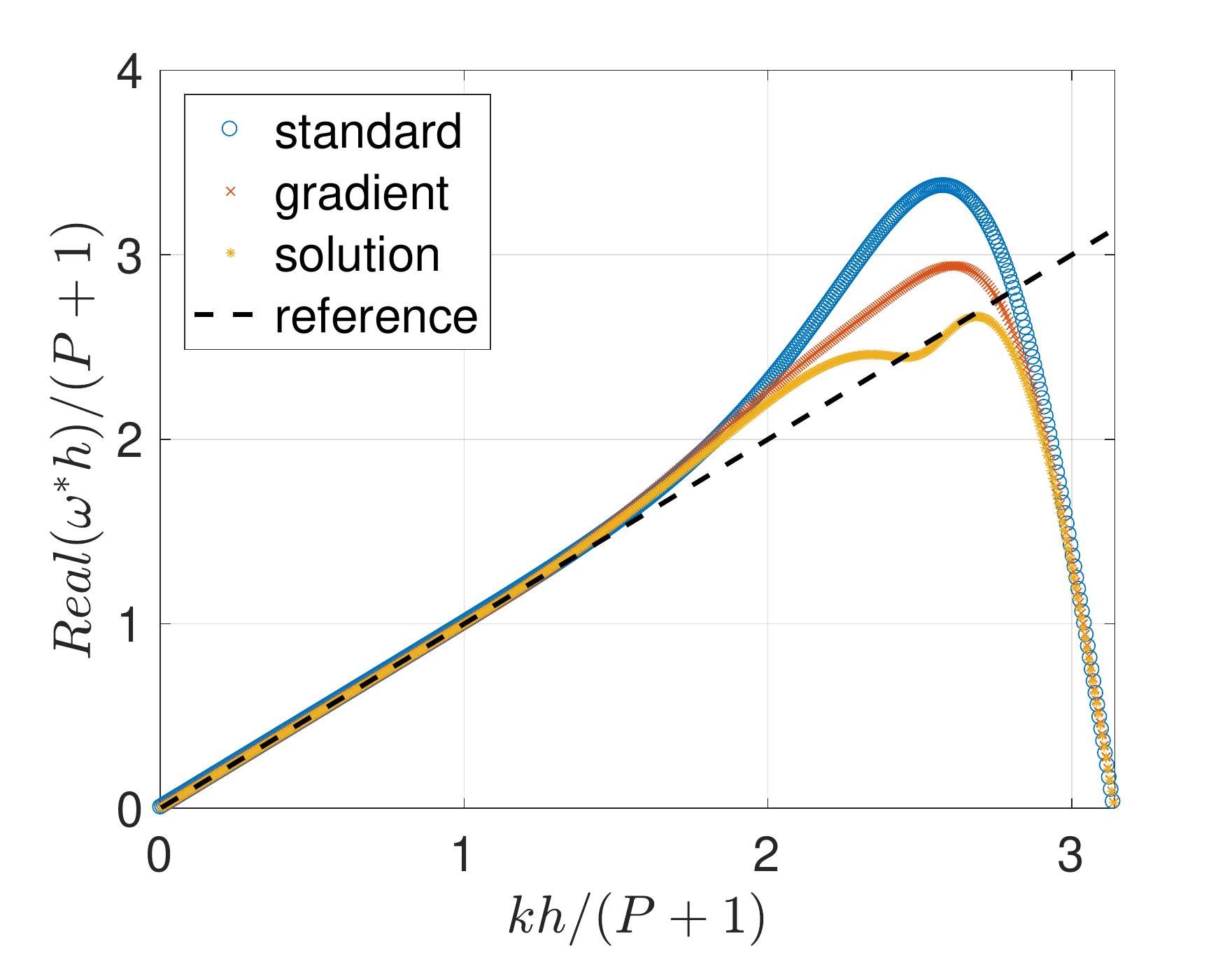}
		\caption{}
	\end{subfigure}
	\begin{subfigure}{.45\textwidth}
		\includegraphics[width=180pt]{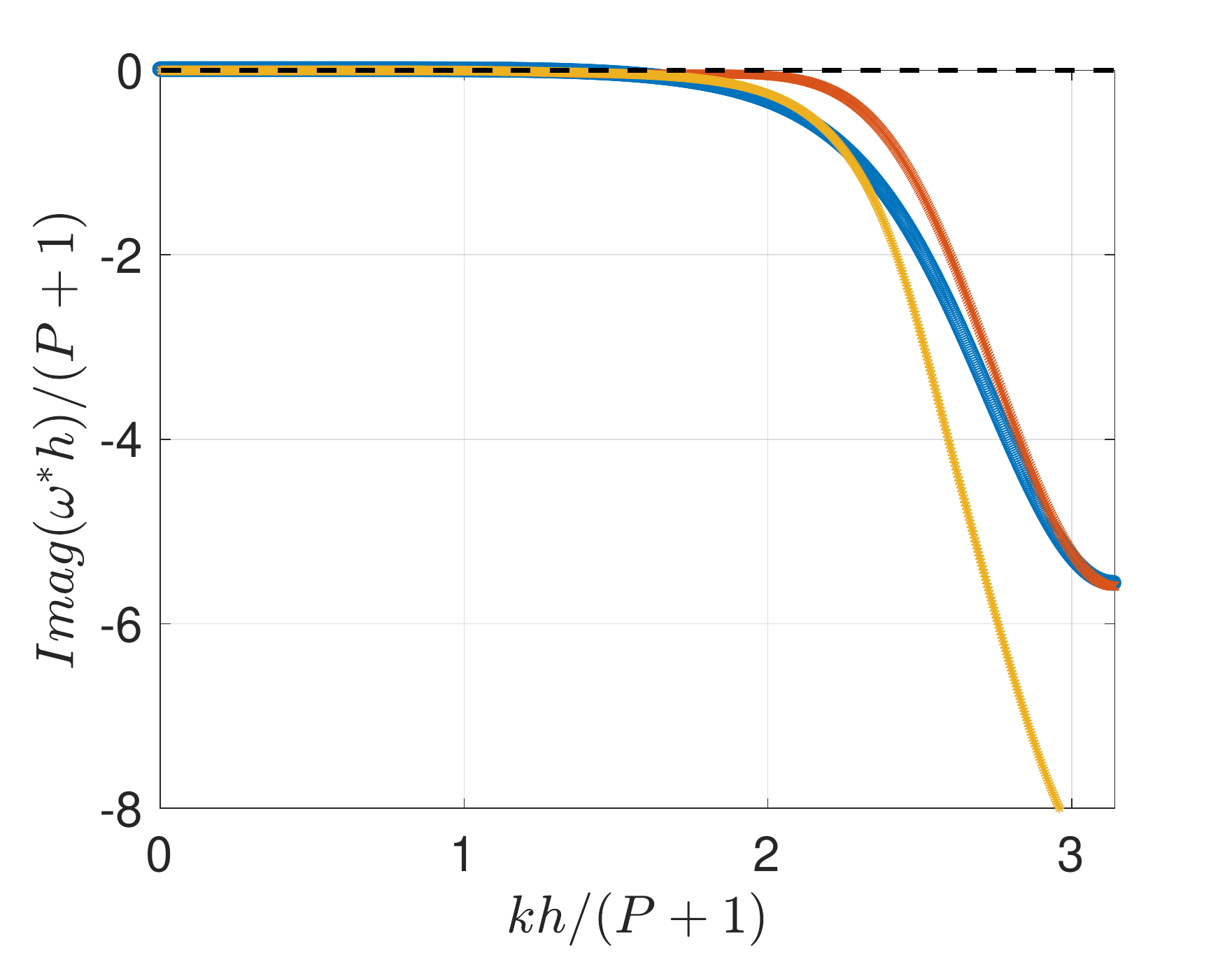}
		\caption{}
	\end{subfigure}
	\centering
	\caption{Optimal dispersion-dissipation behaviour at P = 4 (upwind flux, $\tau_g = -4\times 10^{-4}, \tau_s = 0.25$). a) Dispersion. b) Dissipation.}
	\label{fig:P4-comp}
\end{figure*}

Figures \ref{fig:P2-comp} and \ref{fig:P4-comp} compare the optimal spectral behaviours between two methods of jump penalty stabilisation at polynomial order $P=2$ and $P=4$. Similarly, both the solution and the gradient jump penalty stabilisation methods improve the spectral properties. These results indicate that the proposed jump penalty stabilisation methods can be applied to different polynomial orders. 

\section{Additional results on Burgers turbulence}
\label{sec:appendBT}
\subsection{Other polynomial orders}
\begin{figure*}[htbp]
	\begin{subfigure}{1.0\textwidth}
		\includegraphics[width=1.0\textwidth]{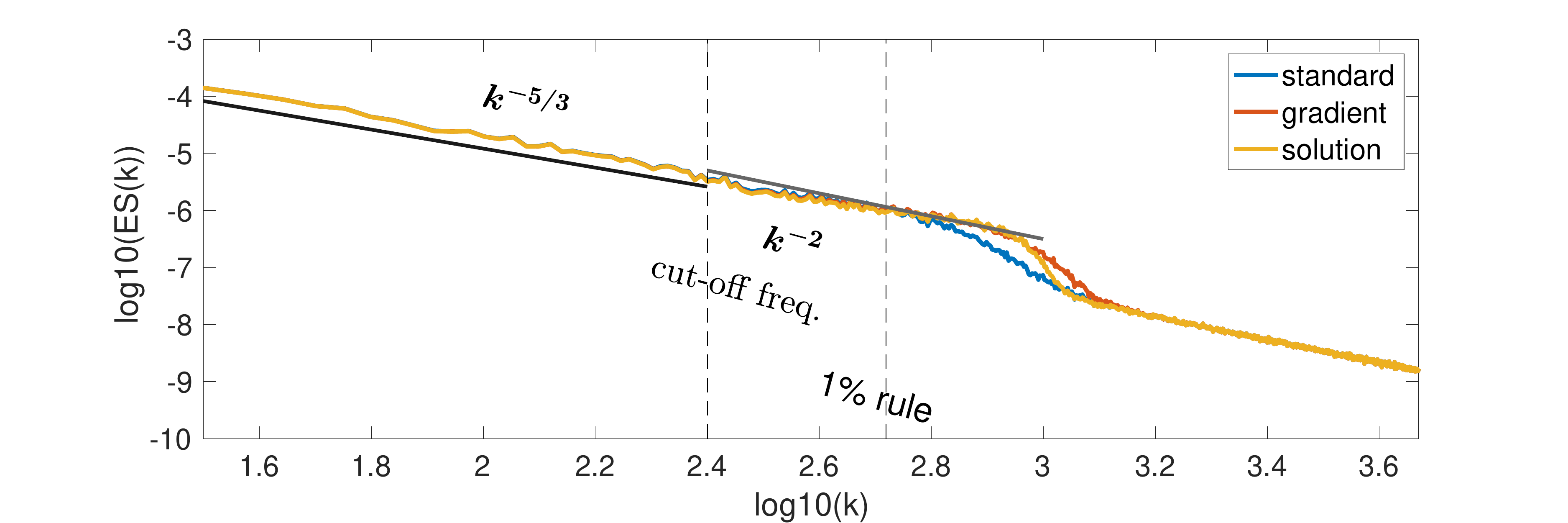}
		\caption{Overview.}
	\end{subfigure}
	\begin{subfigure}{1.0\textwidth}
		\includegraphics[width=1.0\textwidth]{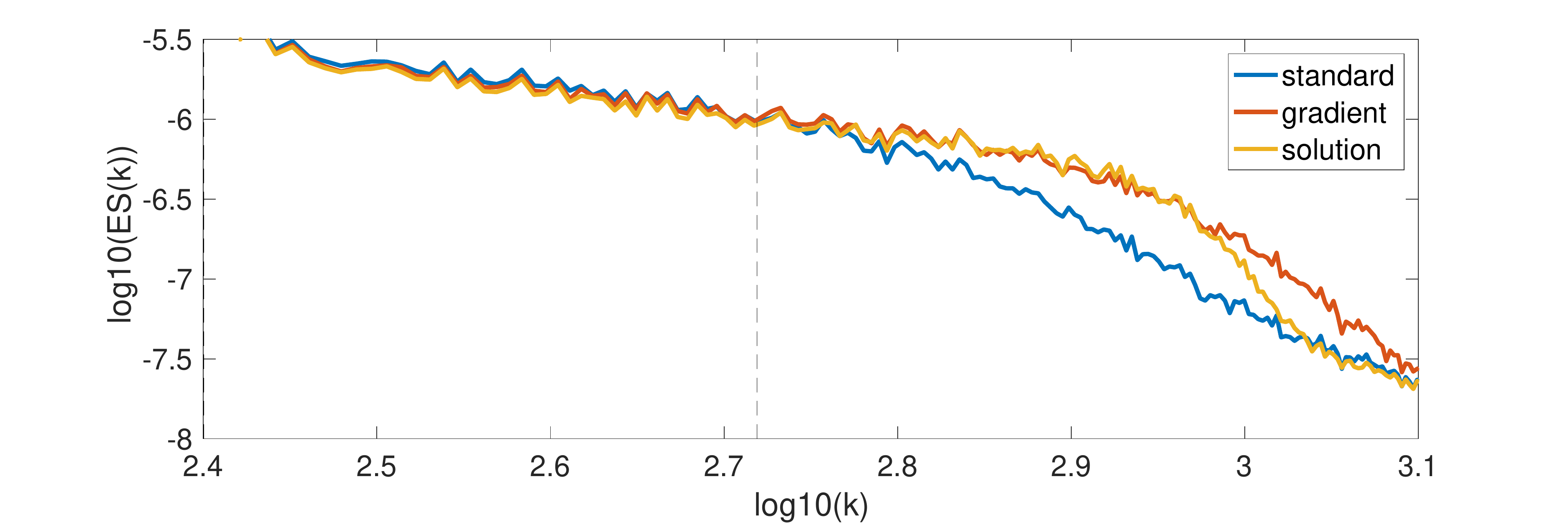}
		\caption{Zoom-in view.}
	\end{subfigure}
	\centering
	\caption{Time-averaged energy spectrum for the forced Burgers turbulence. The polynomial order is $P = 2$ and the number of element is $N = 400$, resulting in $1200$ total DOFs. This simulation is carried out with Roe fluxes, which are similar to the upwind fluxes considered in the advection equation. (Gradient jump penalty: $\tau_g = -2 \times 10^{-3}$; Solution jump penalty: $\tau_s = 1.5$)}
	\label{fig:Burgers-400-P2}
\end{figure*}

\begin{figure*}[htbp]
	\begin{subfigure}{1.0\textwidth}
		\includegraphics[width=1.0\textwidth]{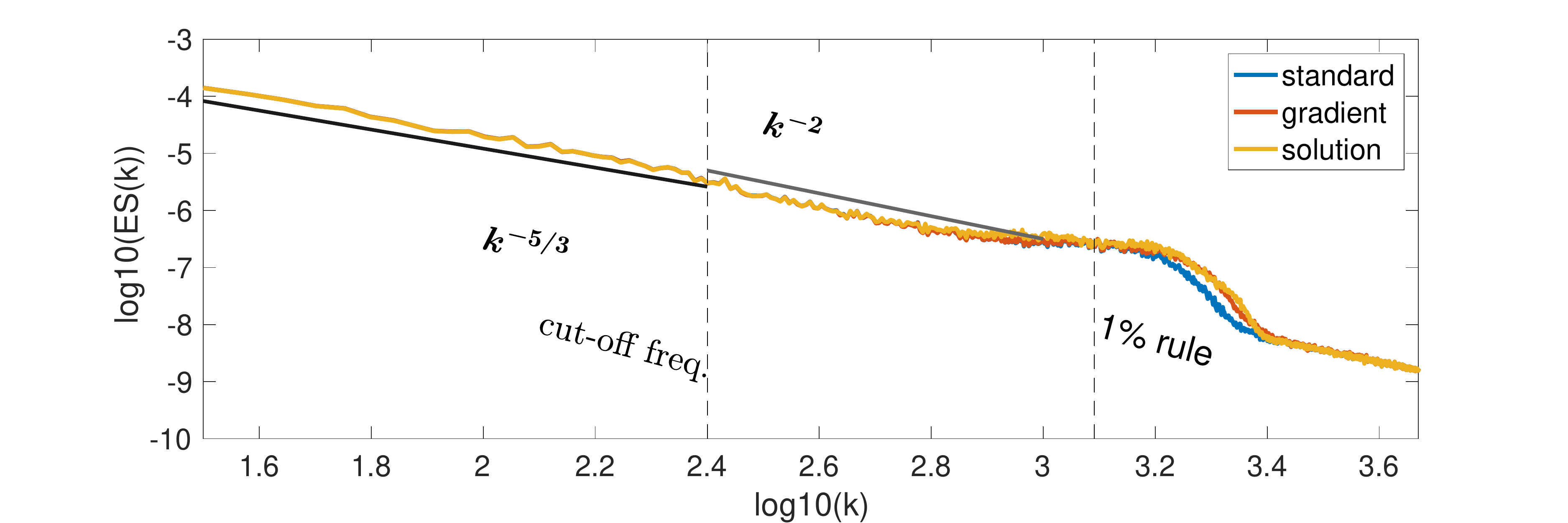}
		\caption{Overview.}
	\end{subfigure}
	\begin{subfigure}{1.0\textwidth}
		\includegraphics[width=1.0\textwidth]{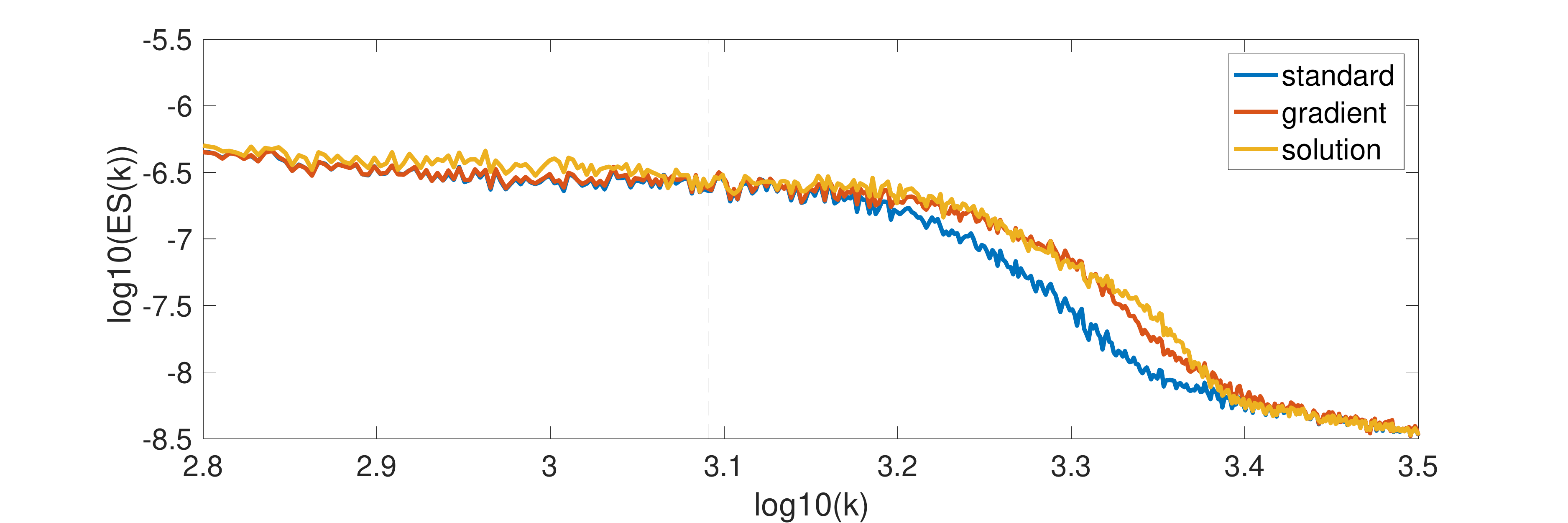}
		\caption{Zoom-in view.}
	\end{subfigure}
	\centering
	\caption{Time-averaged energy spectrum for the forced Burgers turbulence. The polynomial order is $P = 4$ and the number of element is $N = 400$, resulting in $2000$ total DOFs. This simulation is carried out with Roe fluxes, which are similar to the upwind fluxes considered in the advection equation. (Gradient jump penalty: $\tau_g = -3 \times 10^{-4}$; Solution jump penalty: $\tau_s = 1.5$)}
	\label{fig:Burgers-400-P4}
\end{figure*}

Here, the influence of different polynomial orders for $N=400$ has been studied, and the energy spectrum is shown in Figure \ref{fig:Burgers-400-P2} ($P = 2$) and Figure \ref{fig:Burgers-400-P4} ($P = 4$). As the polynomial order varies, both types of jump penalty stabilisation still improve the representation of under-resolved turbulence scales, compared with the standard DG scheme. Similar performance has been observed between gradient and solution jump penalty stabilisations. It is also observed that as the polynomial order increases from $2$ to $4$, the discrepancy between the jump penalty and the standard DG method becomes smaller, since the continuity across the element interface is better maintained at high polynomial orders.

\subsection{Refined resolution}
\begin{figure*}[htbp]
	\begin{subfigure}{1.0\textwidth}
		\includegraphics[width=1.0\textwidth]{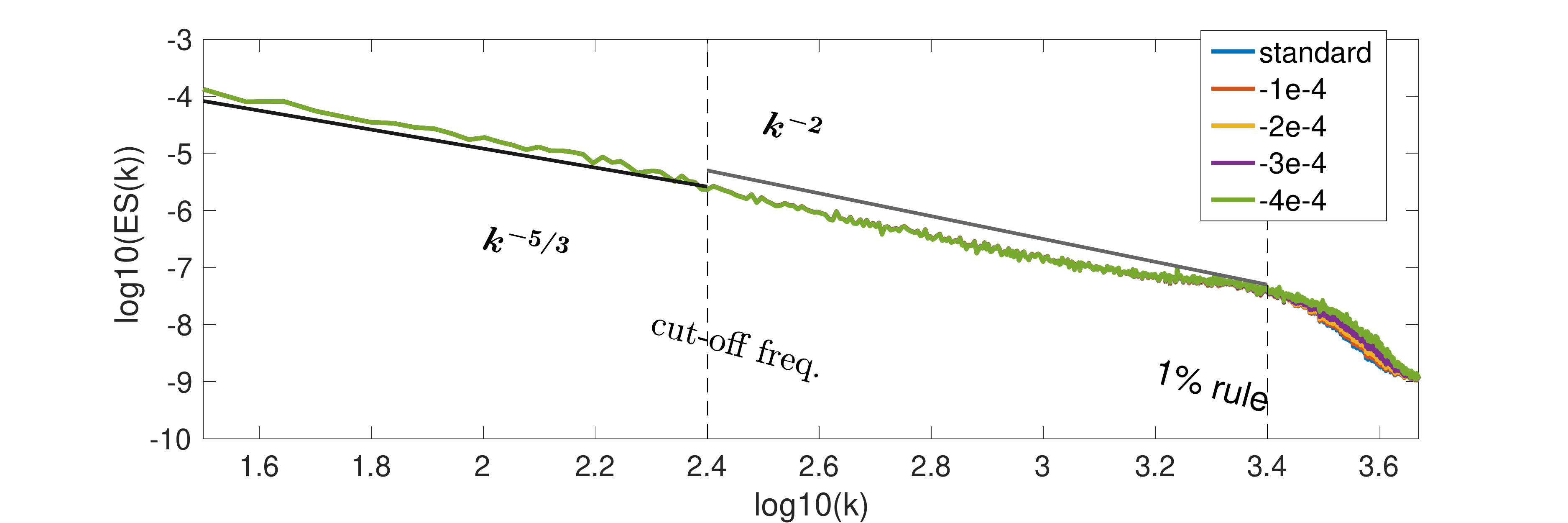}
		\caption{Overview.}
	\end{subfigure}
	\begin{subfigure}{1.0\textwidth}
		\includegraphics[width=1.0\textwidth]{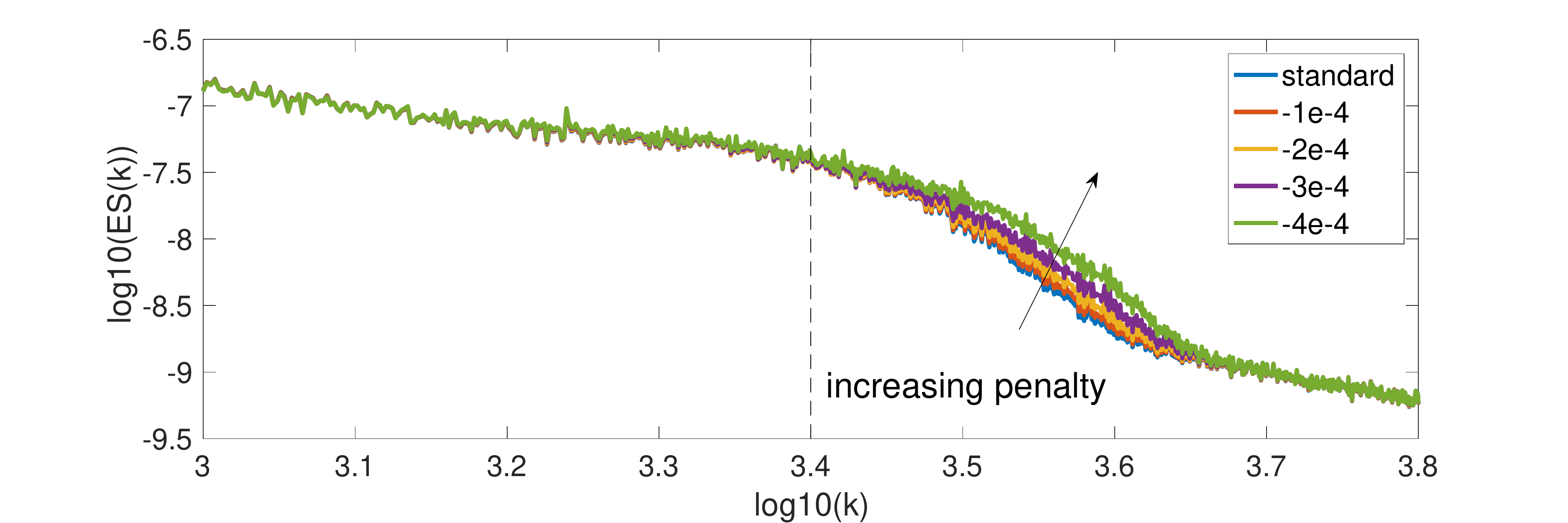}
		\caption{Zoom-in view.}
	\end{subfigure}
	\centering
	\caption{Time-averaged energy spectrum for the forced Burgers turbulence (gradient jump penalty). The polynomial order is $P = 4$ and the number of element is $N = 819$, resulting in $4096$ total DOFs. This simulation is carried out with Roe fluxes, which are similar to the upwind fluxes considered in the advection equation.}
	\label{fig:Burgers-819g}
\end{figure*}

\begin{figure*}[htbp]
	\begin{subfigure}{1.0\textwidth}
		\includegraphics[width=1.0\textwidth]{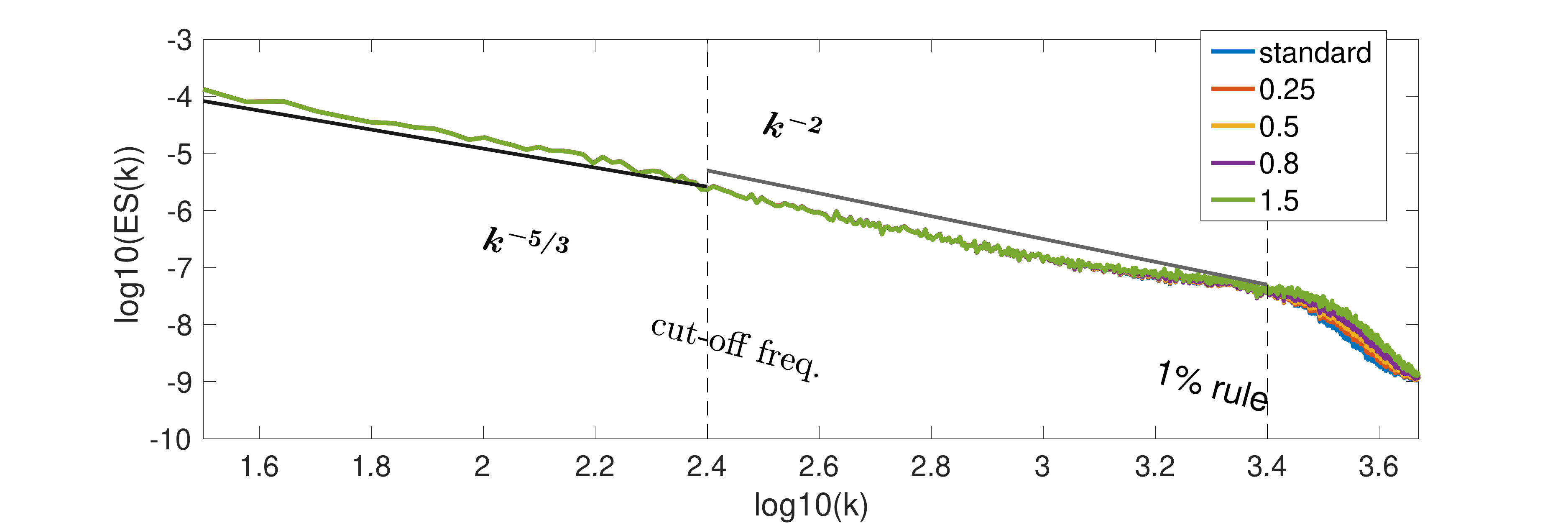}
		\caption{Overview.}
	\end{subfigure}
	\begin{subfigure}{1.0\textwidth}
		\includegraphics[width=1.0\textwidth]{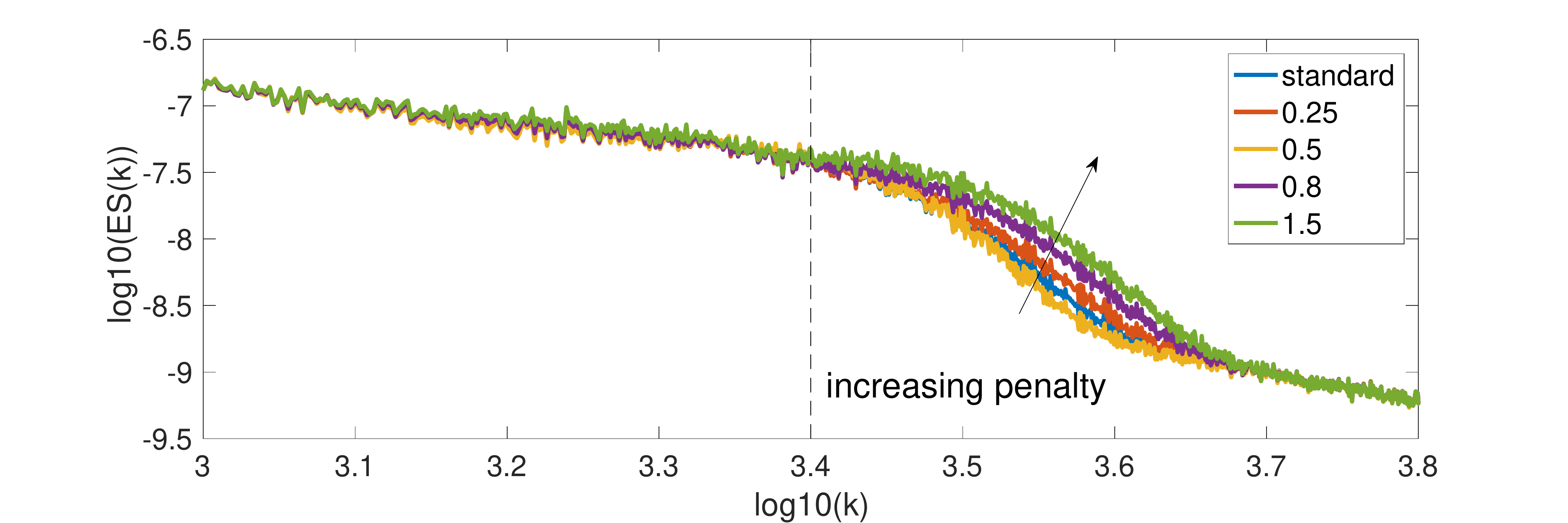}
		\caption{Zoom-in view.}
	\end{subfigure}
	\centering
	\caption{Time-averaged energy spectrum for the forced Burgers turbulence (solution jump penalty). The polynomial order is $P = 4$ and the number of element is $N = 819$, resulting in $4096$ total DOFs. This simulation is carried out with Roe fluxes, which are similar to the upwind fluxes considered in the advection equation.}
	\label{fig:Burgers-819s}
\end{figure*}

This section provides additional results for the Burgers turbulence simulation, considering refined resolution with mesh size $N=819$. The polynomial order is set to $P=4$, resulting in 4096 DOFs. Firstly, parametric studies are presented in Figure \ref{fig:Burgers-819g} (gradient jump penalty stabilisation) and Figure \ref{fig:Burgers-819s} (solution jump penalty stabilisation). Since more DOFs have been used to simulate turbulent flow (4096 DOFs), the resolved wavenumber range becomes larger. As the penalisation term is increased, the wavenumber resolution for under-resolved turbulence simulation is improved, as evidenced by the extended range of the resolved wavenumber range. To determine the optimal penalty parameter, we choose $\tau_g$ or $\tau_s$ that maintains the slope of the forced Burgers turbulence ($-2$) while better resolving the turbulence energy spectrum. Therefore, the optimal penalty parameters for both methods are determined as $\tau_g = -4 \times 10^{-4}$ and $\tau_s = 1.5$, respectively. It should be noted that if the penalisation term continues to increase, the slope of $-2$ cannot be maintained, and thus the real flow physics cannot be described properly. This has been observed in Sec. \ref{sec:case-BT}.

\begin{figure*}[htbp]
	\begin{subfigure}{1.0\textwidth}
		\includegraphics[width=1.0\textwidth]{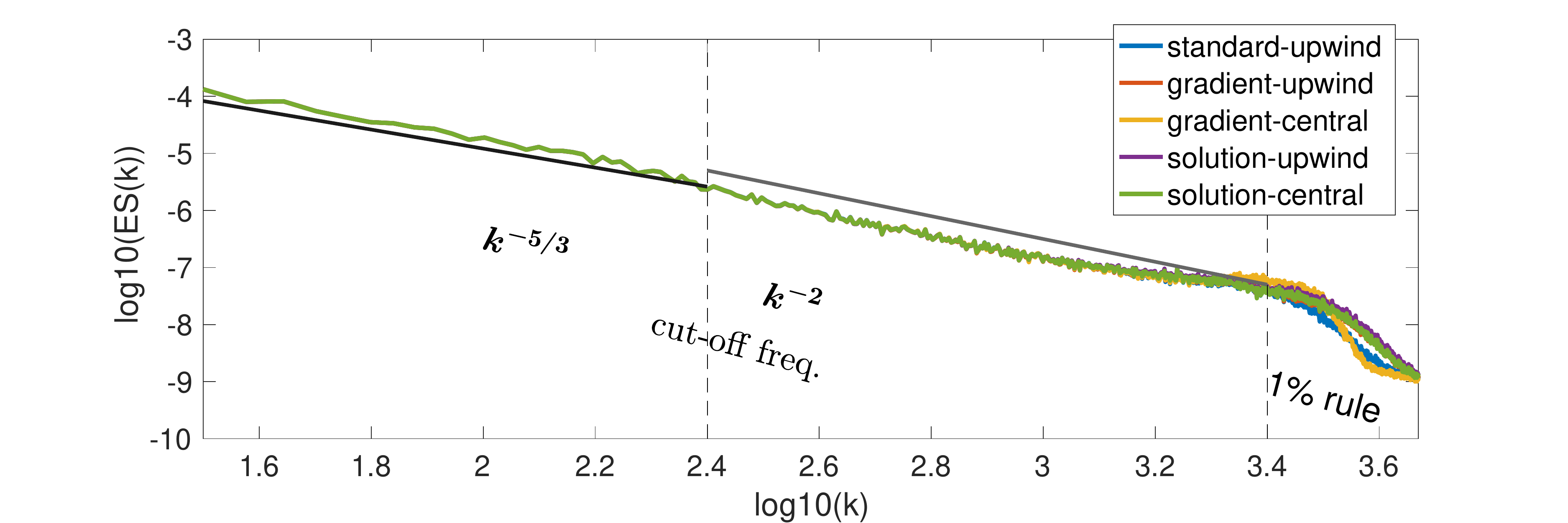}
		\caption{Overview.}
	\end{subfigure}
	\begin{subfigure}{1.0\textwidth}
		\includegraphics[width=1.0\textwidth]{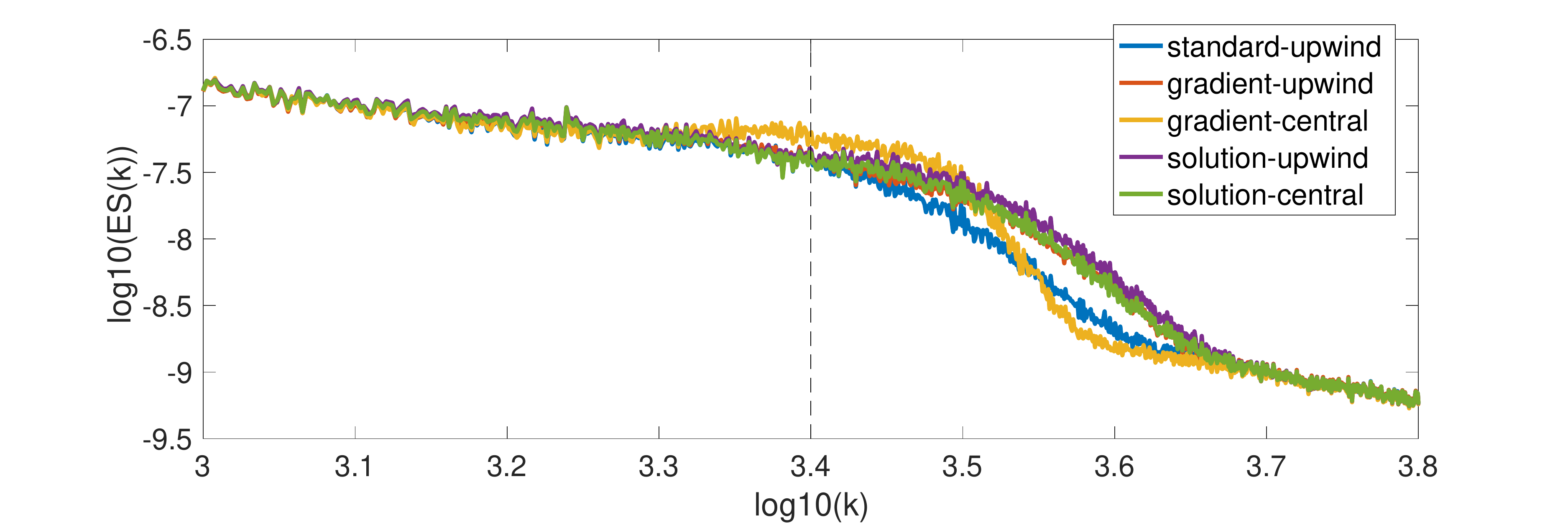}
		\caption{Zoom-in view.}
	\end{subfigure}
	\centering
	\caption{Time-averaged energy spectrum for the forced Burgers turbulence. The polynomial order is $P = 4$ and the number of element is $N = 819$, resulting in $4096$ total DOFs. Different Riemann fluxes have been considered. (Gradient-upwind: $\tau_g = -4 \times 10^{-4}$; Gradient-central: $\tau_g = 0.005$; Solution-upwind: $\tau_s = 1.5$; Solution-central: $\tau_s = 1.5$)}
	\label{fig:Burgers-819}
\end{figure*}

The time-averaged energy spectrum for optimal penalty parameters is shown in Figure \ref{fig:Burgers-819}. Meanwhile, to compare the effect of upwinding on Riemann solvers, both central and upwind fluxes have been considered. It has been found that when the central flux is used, the standard DG simulation will blow up. However, jump penalty stabilisation helps to stabilise the simulation under central fluxes. As shown in Figure \ref{fig:Burgers-819}, for the gradient jump penalty with central flux, although the dissipation drops from a larger wavenumber, the slope of $-2$ is not well maintained. On the contrary, for the rest of the simulations, the resolved range is enlarged, while the expected slope is maintained. The rest of three cases, including solution jump penalty for central flux, and solution and gradient jump penalty for upwind flux, offer similar spectral behaviour under the optimal penalty parameter. These results all indicate that jump penalty stabilisation improves the Burgers turbulence simulation by extending the resolved wavenumber range and providing necessary diffusion in the dissipation range.




\bibliographystyle{elsarticle-num-names}
\bibliography{sample}







\end{document}